%% file: sample-authordraft.tex
\documentclass[manuscript]{acmart}

\usepackage{enumitem}
\usepackage{booktabs}
\usepackage{tabularx}
\usepackage{xcolor,colortbl}
\usepackage{tabularx}
\definecolor{lightgray}{gray}{0.93}
\definecolor{slightgray}{gray}{0.98}
\definecolor{darkgray}{gray}{0.77}
\usepackage{makecell}
\usepackage{multirow}
\usepackage{threeparttable}
\usepackage{hyperref}

\usepackage{datetime2}

\usepackage{graphicx}
\RequirePackage[normalem]{ulem} %DIF PREAMBLE
\RequirePackage{color}\definecolor{RED}{rgb}{1,0,0}\definecolor{BLUE}{rgb}{0,0,1} %DIF PREAMBLE
\newcolumntype{Y}{>{\centering\arraybackslash}X}
\AtBeginDocument{%
  \providecommand\BibTeX{{%
    \normalfont B\kern-0.5em{\scshape i\kern-0.25em b}\kern-0.8em\TeX}}}
\usepackage{xcolor} % 颜色支持
\usepackage{ifthen} % 条件判断

\newboolean{showcomments}
\setboolean{showcomments}{true}

\newcommand{\mycomment}[2][CYN]{%
    \ifthenelse{\boolean{showcomments}}%
        {\textcolor{red}{\textbf{[#1: #2]}}}%
        {}%
}
\newcommand{\tool}{SpecifyUI}
\begin{document}

\title{SpecifyUI: Supporting Iterative UI Design Intent Expression through Structured Specifications and Generative AI}
% \renewcommand{\shorttitle}{ChatScratch}
% \affiliation{%

\author{Yunnong Chen}
\affiliation{%
  \institution{Zhejiang University}
  \city{Hangzhou}
  \country{China}}
\email{chen_yn@zju.edu.cn}

\author{Chengwei Shi}
\affiliation{%
  \institution{Beihang University}
  \city{Beijing}
  \country{China}}
\email{shichengwei@buaa.edu.cn}

\author{Liuqing Chen}
\authornote{Corresponding author.}
\affiliation{%
  \institution{Zhejiang University}
  \city{Hangzhou}
  \country{China}}
\email{chenlq@zju.edu.cn}

\renewcommand{\shortauthors}{Chen et al.}

\begin{abstract}
Large language models (LLMs) promise to accelerate UI design, yet current tools struggle with two fundamentals: externalizing designers’ intent and controlling iterative change. We introduce SPEC, a structured, parameterized, hierarchical intermediate representation that exposes UI elements as controllable parameters. Building on SPEC, we present SpecifyUI, an interactive system that extracts SPEC from UI references via region segmentation and vision–language models, composes UIs across multiple sources, and supports targeted edits at global, regional, and component levels. A multi-agent generator renders SPEC into high-fidelity designs, closing the loop between intent expression and controllable generation. Quantitative experiments show SPEC-based generation more faithfully captures reference intent than prompt-based baselines. In a user study with 16 professional designers, SpecifyUI significantly outperformed Stitch on intent alignment, design quality, controllability, and overall experience in human–AI co-creation. Our results position SPEC as a specification-driven paradigm that shifts LLM-assisted design from one-shot prompting to iterative, collaborative workflows.
\end{abstract}

\keywords{}

\keywords{User Interface Design, Design Intent Expression, Intermediate Representations, Controllable Generation}

\begin{teaserfigure}
  \includegraphics[width=\textwidth]{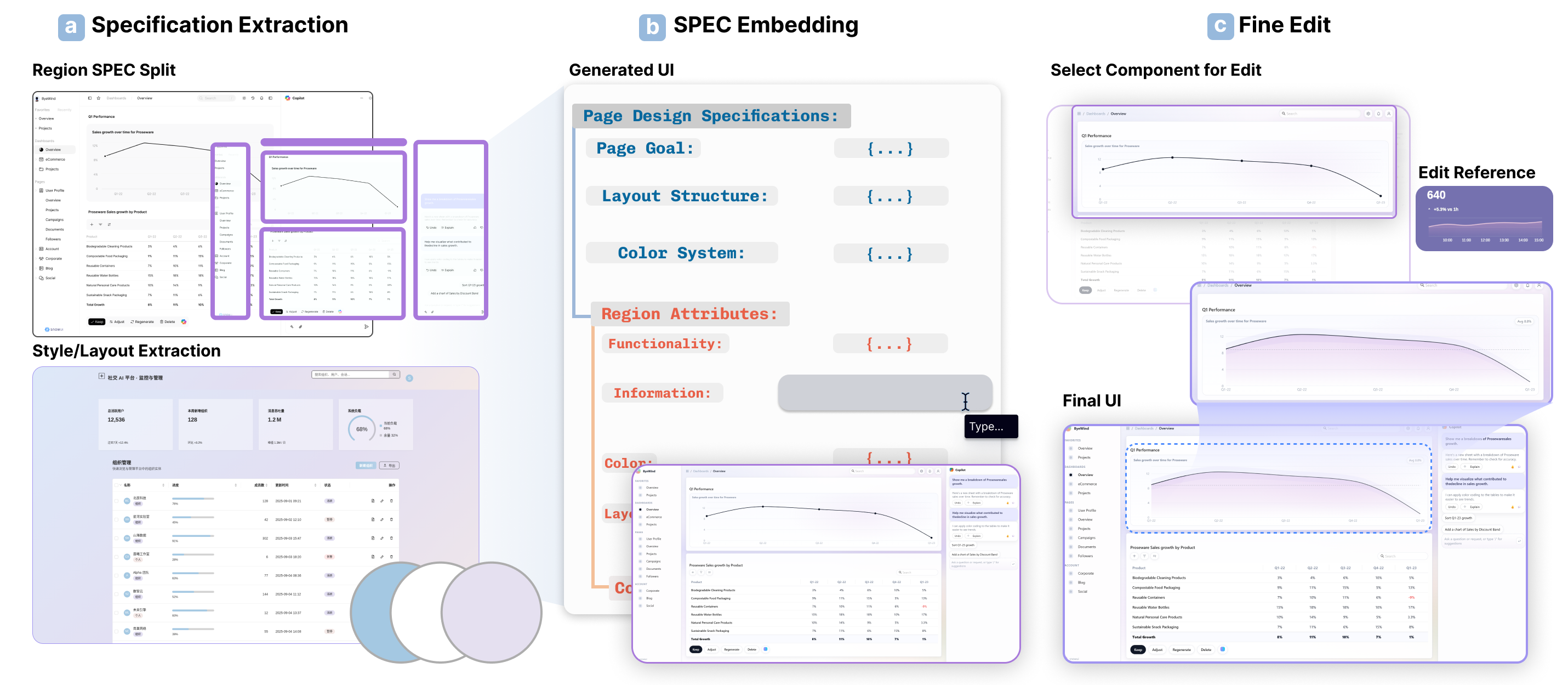}
  \caption{\textbf{SpecifyUI introduces a vision-centered intermediate representation to make design intent explicit and controllable in UI generation: (a) the user begins by extracting specifications, identifying design elements such as Region, Style, and Layout; (b) these elements are then composed into the intermediate representation, SPEC Embedding, to form a coherent whole; (c) the design can be iteratively refined by directly selecting regions and extracting elements, enabling exploration of multiple variants.}}
  \Description{Enjoying the baseball game from the third-base
  seats. Ichiro Suzuki preparing to bat.}
  \label{fig:teaser}
\end{teaserfigure}


\maketitle

\input{sections/Introduction2}
\input{sections/Related_Work}
\input{sections/Formative_Study}
\input{sections/TechnicalPipeline}
\input{sections/SystemOverview}
\input{sections/Evaluation}
\input{sections/Discussion}
\input{sections/Conclusion}

\bibliographystyle{ACM-Reference-Format}
\bibliography{sections/Reference}
\end{document}

%% file: sections/Introduction2.tex
\section{Introduction}
Recent advances in large language models (LLMs) have introduced powerful capabilities for user interface (UI) design, including transforming natural language \cite{wu2024uicoder, ma2025dynex} or visual inputs \cite{gui2025uicopilot, wu2025mllm, wan2025divide, gui2025latcoder} into UI prototypes and even executable code. These abilities reduce repetitive work and accelerate early ideation by rapidly surfacing alternative layouts, styles, and interaction patterns. Beyond ideation, UI design proceeds through iterative refinement. Current LLM-based commercial tools—such as UIzard \cite{uizard} and Stitch \cite{GoogleStitch}—operationalize this by enabling prompt-based edits to generated prototypes, supporting exploration across versions, and facilitating collaborative review. Taken together, automated generation coupled with prompt-mediated refinement signals an emerging LLM-centric design paradigm with the potential to transform UI workflows.

Despite these advances, current LLM-based methods face two major challenges. The first challenge is intent expression. UI design intent is often abstract and multifaceted, making it difficult for LLMs to accurately interpret \cite{lu2025misty, shokrizadeh2025dancing}. Without designers’ deliberate planning, LLMs struggle to produce professional designs that effectively convey intended visual messages \cite{park2025leveraging,kim2025interacting}. A key barrier is communication \cite{subramonyam2024bridging}: LLMs operate primarily through textual instructions, whereas designers predominantly think and communicate in visual terms \cite{weisz2024design}, leading to outputs that frequently miss stylistic direction or misrepresent hierarchical relationships. The second challenge is controllability. UI design is inherently iterative, requiring designers to refine ideas across global, regional, and component levels. Yet current generative design tools \cite{GenUIStudy, chen2024autospark} provide limited means for scoped edits and often lack predictable behaviors: due to the stochastic nature of LLMs, modifications can unintentionally alter unrelated parts of the design, undermining consistency across iterations.

A promising direction to address these challenges is the use of intermediate representations \cite{jiang2024self, li2025structured}, which can balance semantic and controllability in LLM-assisted UI generation. Existing approaches generally fall into two categories: domain-specific languages (DSLs) and semi-formal specifications. DSLs offer precise control by enforcing rigid syntactic rules \cite{rocha2022towards, mernik2005and}, but they are often too restrictive and inaccessible for design tasks, as they require designers to author code-like expressions rather than work in familiar design workflows. Semi-formal specifications, by contrast, incorporate semantic cues to guide generation \cite{cao2025generative}, but existing methods \cite{cao2025generative, ma2025dynex} have mainly focused on producing UI prototypes from high-level intent, overlooking hierarchical structure and design guidelines that are central to professional design practice.

Given this mismatch, we conducted a formative study with six professional UI designers. The study confirmed that intent expression and controllability remain critical challenges in practice. It further revealed three pain points specific to UI design: (i) designers struggle to describe visual elements such as layout balance, style, and hierarchy using natural language; (ii) they find it difficult to articulate how these elements should be composed into a coherent interface; and (iii) iterative refinement is inefficient, as designers must manually track what to change and what to keep in long prompts, which is tedious and error-prone.

To address these challenges, we introduce SPEC, a structured, parameterized, and hierarchical intermediate representation that encodes UI design guidelines and element hierarchies, aligning user intent with LLM-based generation. Built on SPEC, we designed an LLM-driven design tool that leverages region segmentation models and vision–language models (VLMs) to extract SPEC directly from UI references, enabling designers to express visual elements more naturally. With SPEC, designers can flexibly combine elements extracted from multiple references, while VLMs ensure that the resulting representation remains stylistically and structurally coherent. Finally, to support iterative exploration, SPEC allows designers to directly target modifications at global, regional, or component levels, ensuring that multi-round refinements remain structure-preserving. We further propose a multi-agent collaborative UI generator that faithfully translates SPEC into high-fidelity UI designs, closing the loop between intent externalization and controllable generation.

We systematically evaluated SPEC and SpecifyUI through quantitative experiments and a user study. The quantitative results demonstrated that SPEC-based generation more faithfully captured reference design intent compared to prompt-based baselines. In a user study with 16 designers, our system was compared against Stitch, an LLM-driven commercial design tool with conversational interaction, and was rated significantly higher in clarity of intent expression, controllability of generation, and overall user experience. These findings demonstrate that SPEC not only improves the fidelity of generated UIs but also shifts the paradigm of LLM-assisted design: from prompt-driven, one-shot generation toward specification-driven, iterative collaboration, where designer intent is explicitly externalized and preserved across refinements.

Therefore, this work makes the following contributions:

\begin{itemize}
    \item \textbf{SPEC}: an intermediate representation of UIs that is structured and parameterized, encoding layout, style, and content into controllable parameters.
    \item \textbf{Pipeline}: a UI generation pipeline that constructs and edits SPECs, leveraging a multi-agent framework to produce and refine UIs with improved fidelity and controllability.
    \item \textbf{System}: \tool, an interactive system that enables designers to compose SPECs, generate UIs, and refine designs through controllable, structure-preserving edits.
    \item \textbf{Evaluation}: a comprehensive assessment combining technical benchmarks against prompt-based baselines and a user study with 16 designers against a commercial baseline, showing significant improvements in intent expression, controllability, and user experience.
\end{itemize}

%% file: sections/Related_Work.tex
\section{Related Work}

\subsection{Intermediate Representations for UI Generation}
Intermediate representation (IR)–driven UI generation has long been central to UI engineering, aiming to bridge the semantic gap between high-level design intent and concrete implementation \cite{jacob1983using, myers2000past, hudson1990interactive}. Early Model-Based User Interface (MBUI) research introduced formal specifications to enhance rigor and reduce development effort, typically through abstract task models, domain models, and mapping rules that translate specifications into UI components. For example, Just-UI \cite{molina2002just} and the Auckland Layout Editor \cite{zeidler2013auckland} provided modeling environments, while Nichols et al. mapped functional requirements directly into UI specifications \cite{nichols2002generating, nichols2004improving, nichols2006uniform}. Hudson et al. \cite{hudson1994user} even employed spreadsheet metaphors for partially automated layouts. Despite these contributions, MBUI approaches were largely developer-oriented and static, limiting adaptability and end-user involvement.

Building on this foundation, more recent systems sought to make specifications more accessible and adaptable, empowering end-users to generate or modify UIs. Bespoke \cite{vaithilingam2019bespoke} maps command-line parameters to widgets, DynaVis \cite{vaithilingam2024dynavis} employs Vega-Lite for visualization editing, and NL4DV \cite{narechania2020nl4dv} translates natural language into visualization specifications. More recent systems integrate specifications with generative workflows: DynEx \cite{ma2025dynex} treats specifications as modular, stepwise implementation plans that prioritize procedural control and version management, while Jelly \cite{cao2025generative} builds evolving task-driven models from natural language and direct manipulation to support malleable UIs. These efforts demonstrate the value of constraints and scaffolds but still provide limited support for multi-layered design semantics. Importantly, neither MBUI nor specification-based systems have been integrated with LLM-driven UI generation, where ambiguity in intent expression and lack of controllability remain major barriers. Our work addresses this by introducing SPEC, a structured, semantically expressive IR designed for LLM integration and co-creative UI design.

\subsection{AI Support for UI Design Generation}
Recent advances in deep generative models and LLMs have enabled automatic UI generation from visual or textual inputs, supporting early-stage ideation \cite{gajjar2021akin, jiang2024graph4gui, wu2024uicoder}. GAN-based methods such as GUIGAN \cite{zhao2021guigan} recombine components, and GanSpiration \cite{mozaffari2022ganspiration} blends UI styles with unrelated images to inspire new designs. Layout-centric approaches, including Layout Transformer \cite{gupta2021layouttransformer}, LayoutVQ-VAE \cite{jing2023layout}, LayoutFlow \cite{10.1007/978-3-031-72764-1_4}, and Spot the Error \cite{lin2024spot}, capture semantic and spatial relationships to produce coherent structures. These approaches demonstrate the promise of generative models for creating structural and stylistic variation.

While these deep generative approaches reveal the potential of model-driven UI creation, another line of work employs LLMs for code generation, translating natural language into HTML or React \cite{gui2025uicopilot, chen2025designcoder, wan2025divide}. UICopilot \cite{gui2025uicopilot} synthesizes webpages through hierarchical code generation, while LayoutCoder \cite{wu2025mllm} leverages multimodal models and layout-aware decoding to generate code from webpage images. Commercial tools, including Uizard \cite{uizard}, Galileo \cite{GalileoAI}, and Google Stitch \cite{GoogleStitch}, as well as AI-enhanced features in Figma and Framer, extend these ideas in practice. For example, Stitch employs Gemini to transform prompts or screenshots into HTML prototypes. Despite their power, such systems are prompt-driven, meaning small textual changes can disrupt entire designs, continuity across iterations is poor, and designers lack direct control over layout or semantics. These limitations underscore a critical gap: existing generative approaches fall short of interactive design workflows, where continuity, controllability, and clear intent expression are essential. To address this, we propose leveraging structured UI specifications as an intermediate layer, enabling designers to flexibly edit layouts, components, and semantics while ensuring that modifications are faithfully reflected in outputs—aligning generation with human design practices.

\subsection{Specifying Design Intent through Human–AI Collaboration}
% 通过交互策略指定意图
% 仅用自然语言指定设计意图的挑战已得到充分认识[13,80,85,98]。Subramonyam 等[83]对用户如何将其目标转化为明确的意图进行了理论化，强调了生成模型对语言精度高度敏感，但人类语言容忍表达变体以传达相似含义的指令差距。这种挑战在各个设计领域的研究中都很明显，其中临时用户[ 60]、平面设计师[ 55]、制造设计师[ 35]和游戏专业人士[ 90]努力通过繁琐的即时工程过程来表达视觉意图。这种将视觉设计转化为语言媒介的摩擦，可以通过设计方法论文献的视角来理解[23,86,95]。最近促进意图表达的努力分为三类：1）将冗长的提示分解为模块化提示;2） 用其他模式增强文本提示;3）通过直接作解决文本提示中的歧义[79]。
% 模块化提示。管理冗长的提示具有挑战性，许多研究已经探索将它们模块化为可管理的部分。例如，人工智能链[ 100]支持链接单个提示以进行端到端执行。本着类似的精神，ChainForge [ 4] 支持比较模型之间的提示变化。在视觉设计中，Keyframer [ 88] 使用“分解提示”进行分步动画设计。同样，Spellburst [ 3] 和 ComfyUI [ 22] 采用基于节点的接口来模块化创意编码和图像生成，集成了不同的控制信号。尽管这些工具有助于灵活编写和修改提示并允许查看中间结果，但它们仍然需要用户在文本中指定他们的视觉意图，这并不能完全解决意图规范的精确性。
% 多模态提示。最近的商业化工具（例如，DALL ·E3 [ 24]、Adobe Firefly [ 31]、MidJourney [ 62]、Flux AI [ 2]）允许用户上传图像作为全局内容和/或样式参考;但是，用户不能在图像上添加本地注释以进一步说明他们的细粒度意图。Kaiber Superstudio [ 84] 通过允许用户指定单个本地主题来支持字符一致性生成，但不支持多个主题。Krea.ai [ 54] 使用户能够在自己的资产上进行训练，但每个主题需要多个实例。研究工作还调查了多模态提示的不同策略。例如，DesignPrompt [ 68] 允许用户输入文本、图像和颜色，然后帮助将它们翻译成最终的文本提示。然而，将视觉效果转换为文本通常会导致信息丢失，从而难以捕捉精确的视觉身份。PromptCharm [ 94] 使用户能够检查生成图像的哪一部分对应于文本提示的哪一部分。Sarukkai 等[ 74]介绍了从粗到细的草图引导图像生成。尽管这些工具在某种程度上有助于阐明要使用的视觉元素，但它们无法帮助用户表达多个视觉元素如何相互关联，因为它们通常单独处理每个输入模式。
% 通过直接作进行提示。一些工具采用视觉隐喻来帮助用户通过直接作视觉对象而不是文本提示来指定他们的意图。例如，TaleBrush [ 20] 使用草图线来指示叙述过渡，而 PromptPaint [ 19] 为语义提示插值提供了一个类似绘画的界面。手势用于表示编辑意图，例如，绘制蒙版以修复 [ 6]，添加彩色笔触以重新着色 [ 103]，拖动点以编辑姿势或面部表情 [ 67]。虽然直观，但这些方法是特定于任务的，需要用户重新学习每个用例的交互。最近的工作通过旨在将图形用户界面 （GUI） 与自然语言界面 （NLI） 集成来扩展这一方向。DirectGPT [ 61] 允许用户将图形元素拖放到提示上，但这些元素充当孤立的符号，失去了它们的空间关系。DynaVis [ 89] 将 NLI 与动态生成的 GUI 小部件相结合，用于可视化创作，但要求用户在文本中指定编辑意图。DirectGPT 和 DynaVis 都解决了文本提示中连续数字表达式的模糊性，但它们仍然迫使用户思考文本并主要处理文本。
In UI generation, a central challenge is how designers can effectively convey design intent. Natural language prompts are the most common interface, yet prior work has shown that they often fail to capture the nuance of visual goals \cite{simkute2025ironies,tankelevitch2024metacognitive}. Subramonyam et al. \cite{subramonyam2024bridging} theorized how users translate goals into explicit instructions, highlighting a mismatch: generative models demand linguistic precision, whereas human language tolerates ambiguity and variation. 

To address this tension, researchers have explored three directions. First, modularized prompts seek to break down complex specifications into manageable units. AI Chains \cite{wu2022ai} and ChainForge \cite{arawjo2023chainforge} allow chaining or comparing prompts. Node-based interfaces such as Waitgpt \cite{xie2024waitgpt} and Spellburst \cite{angert2023spellburst} integrate multiple control signals to modularize creative coding. These approaches improve flexibility and mid-process inspection, but still require designers to express visual semantics through text, which rarely captures layout or aesthetic constraints.

Second, multimodal prompts extend expressivity by combining text with other inputs. Commercial systems such as Uizard \cite{uizard}, Stitch \cite{GoogleStitch}, and MasterGo \cite{Mastergo} allow designers to upload images or keywords as global style references, but provide little support for localized annotations or structural relationships between components. Research prototypes have begun to enrich multimodality: DesignPrompt \cite{10.1145/3643834.3661588} translates text, images, and colors into prompts, PromptCharm \cite{10.1145/3613904.3642803} visualizes mappings between text and image regions, and sketch-based systems \cite{10.1145/3654777.3676444} enable coarse-to-fine refinement. Nevertheless, these methods often process modalities in isolation, limiting their ability to capture how elements interact within a UI layout.

Finally, direct manipulation approaches bypass text altogether by letting users interact directly with visual objects. TaleBrush \cite{chung2022talebrush} uses sketch lines to denote narrative transitions, while DirectGPT \cite{DirectGPT} allows elements to be dragged into prompts. Recent systems combine GUIs with NLIs: DynaVis \cite{vaithilingam2024dynavis} dynamically generates GUI widgets to complement text, Dynex explores design directions through interactive updates, and Misty \cite{lu2025misty} enables drag-and-drop references to blend multiple UIs. While these techniques reduce reliance on textual encoding, they often treat GUI elements as isolated symbols, overlooking the spatial and hierarchical relationships central to UI design.

%% file: sections/Formative_Study.tex
\section{Formative Study}

\subsection{Method}

To better understand the limitations that designers encounter when applying generative AI in UI practice, we conducted a two-hour formative study with six participants: two UI directors (P1, P6), two UI engineers (P2, P3), and two UI designers (P4, P5). Participants ranged in age from 26 to 40 (\textit{M}=31.5) and all reported more than three years of experience using AI-powered tools. All participants were recruited through industry connections, provided informed consent, and received \$40 compensation for their time.

\begin{table}[ht]
\centering
\caption{Demographic information of participants in the formative study, arranged by age.}
\vspace{-0.11in}
\label{tab:formative_participants}
\begin{tabular*}{0.85\textwidth}{@{\extracolsep{\fill}} c c c c c}
\hline
\textbf{ID} & \textbf{Gender} & \textbf{Age} & \textbf{Job Title} & \textbf{Year of Experience} \\
\hline
P1 & Male & 40  & UI directors & $>$10 years \\
P2 & Male & 35  & UI engineers & $>$10 years \\
P3 & Male & 32 & UI engineers & 5--10 years \\
P4 & Female & 26  & UI designers  & 1--3 years \\
P5 & Female & 27  & UI designers    & 5--10 years  \\
P6 & Female & 29  & UI directors  & 5--10 years \\
\hline
\end{tabular*}
\end{table}

The study consisted of a co-creation task followed by a semi-structured interview. In the co-creation task, participants were asked to redesign a recent UI project using \textit{Stitch}, a generative design tool developed by Google and powered by the Gemini-2.5 model. We selected Stitch because it exemplifies a new class of AI tools that support end-to-end interface generation through natural language prompting. Unlike earlier AI-assisted tools that offered only partial assistance (e.g., color suggestions, layout templates), Stitch enables designers to specify, iterate, and refine UIs entirely via textual input, reflecting a broader shift toward prompt-centric design workflows.  

A 10-minute tutorial introduced Stitch’s main features, followed by a 45-minute design session. Participants shared their screens and used a think-aloud protocol, which allowed us to observe their interactions, strategies, and immediate reactions to system outputs. The session concluded with semi-structured interviews to capture reflections on the utility and challenges of generative UI tools in professional workflows.  

\subsection{Findings}

\textbf{C1. Difficulty in translating visual references into an intermediate representation.}
% 强调视觉转换的重要性，为系统视觉提取和组合作铺垫
Before starting a project, participants routinely collected and arranged screenshots in tools like Figma or Photoshop, grouping them to compare layouts, styles, and spacing (P1, P2, P4–P6). Previous work showing that designers construct external visual structures to reason about references \cite{holinaty2021supporting,koch2020imagesense}, we observed that this practice effectively served as an informal intermediate representation, where participants decomposed and recombined visual features to guide their design decisions. However, current AI tools offer no structured way to carry this visual reasoning forward. Some participants (P2, P5) tried to verbalize details in long prompts, while others (P1, P3, P4, P6) uploaded screenshots, expecting the model to capture salient features. Both strategies revealed the absence of a true visual-to-structural bridge: prompts forced unnatural linguistic translation of perceptual cues, while image inputs often missed or distorted intended elements. As P1 noted, “I want the layout and visual feel from this image, but describing all that in words is too much work.” The difficulty increased even further when merging multiple references (e.g., layout from one, colors from another), as models lacked mechanisms for selective extraction and recombination. Overall, designers’ workflows were grounded in active visual thinking, but LLMs lacked an equivalent intermediate layer to interpret and manipulate visual intent.

\textbf{C2. Difficulty articulating complex visual relationships when composing elements.}
% 强调层级组织的重要性，为系统的层级组合排列功能铺垫
Participants emphasized that UI design involves more than placing components; it requires orchestrating higher-level visual relationships—establishing focal points, balancing hierarchical layers, and maintaining stylistic coherence. These concerns became especially salient when combining multiple reference images, where designers wanted to merge the layout of one with the layout or color scheme of another. Yet current AI tools provided only textual prompts, which proved insufficient for conveying such multi-dimensional relationships. Several participants described this difficulty directly. P3 explained, “I can tell you I want a grid or a sidebar, but it’s very hard to explain how the hierarchy should feel or which part should stand out first.” Similarly, P5 noted, “When I combine two references, I know in my head how they should balance, but I don’t have the words to explain that balance to the AI.” Lacking expressive means for visual reasoning, participants often reduced their prompts to simple component-level instructions (e.g., “add a card,” “use a grid layout”). This forced simplification produced outputs that felt generic and aesthetically weaker, diverging from their intended design vision.

\textbf{C3. Inefficiency in iterating through manual modifications of textual prompts.}
% 强调修改的时候缺少中间表达维护会导致问题
Participants (P2, P5, P6) described iterative refinement with current AI tools as cumbersome and disruptive. Although their natural workflow followed a coarse-to-fine rhythm, each small change in current AI design tools felt like starting over. As P2 explained, “Every change feels like a restart, because the AI doesn’t remember what I wanted before.” The linear, turn-by-turn interaction lacked mechanisms for retaining key design information across iterations, forcing designers to re-describe prior constraints in every prompt. P3 further emphasized the burden of manually preserving context: “I have to restate the entire UI description in each edit just to keep the parts I liked—but even then, the AI often changes them.” This lack of persistent state meant designers had to act as caretakers of the design specification themselves, duplicating large portions of text simply to prevent unwanted changes. Such manual management not only slowed iteration but also interrupted creative flow, making it difficult for participants to progressively refine their ideas.

\subsection{Design Goals}

To address the challenges identified above, we propose the following design goals to guide the development of the \textbf{Specify} system:

\textbf{DG1: Develop effective, structured, and parameterized representations to align LLMs with design intent.} 
Designers typically focus on specific elements within reference images, but current systems require uploading entire images and verbally explaining their emphasis (C1). This creates a gap between what designers attend to and what the system processes. To bridge this gap, the system should allow designers to externalize their selective focus as visual input and capture design intent through structured and parameterized attributes aligned with LLMs. The system should support selecting elements at multiple levels of granularity (screen, region, component) and map them to dimensions such as layout, style, or content. 

\textbf{DG2: Multi-dimensional referencing for compositional relationships.}  
Designers view UI elements as interdependent, yet articulating relationships such as proportion, hierarchy, and spatial proximity remains challenging (C2). To address this, the system should provide an intuitive hierarchical UI structure, where designers can visually arrange and define relationships rather than relying solely on verbal descriptions. The goal is to establish a shared visual structure that allows designers to clearly define compositions while enabling AI to accurately interpret and preserve these complex relationships.  

\textbf{DG3: Facilitate controllable iteration and refinement.}  
Designers found the linear nature of current text-to-UI tools inconvenient, as these systems lack mechanisms for selectively modifying or refining elements across iterations (C3). To address this, the system should support scoped editing, where changes are applied as parameter updates within a structured specification, ensuring that unaffected attributes remain stable. This enables designers to perform coarse-to-fine refinements while preserving design consistency, ultimately aligning AI-assisted workflows with professional practices of non-destructive, iterative design.  

%% file: sections/TechnicalPipeline.tex
\section{Technical Pipeline for Spec-based UI Generation}

\subsection{UI SPEC Schema}

\begin{figure}[t]
\centering
\includegraphics[width=0.99\textwidth]{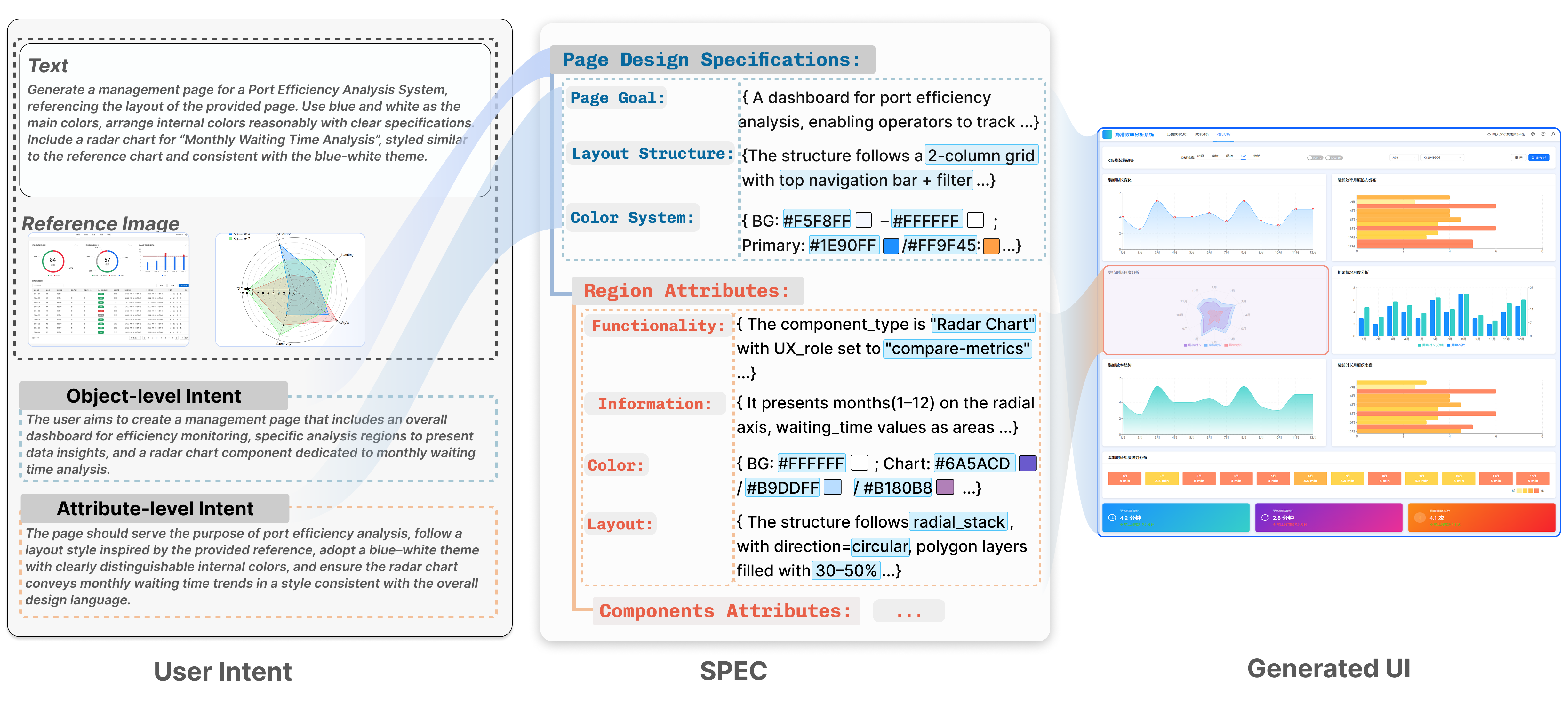}
\vspace{-0.15in}
\caption{\textbf{Mapping user intent into structured UI representations via SPEC. Designers express high-level product goals, page-level composition, and stylistic preferences, which are translated into a multi-layered SPEC representation. This structured prompt then guides large language models to generate UI layouts aligned with the designer’s goals.}}
\label{fig:spec}
\vspace{-0.1in}
\end{figure}

As shown in Figure \ref{fig:spec}, SPEC connects intent and generated UI, illustrating how user intents are mapped into parameterized and semantic attributes that guide the final interface rendering.  
In our modeling, \textbf{SPEC} is formalized as a two-level structural system, consisting of the \textit{Global UI Specification} and the \textit{Page Composition}. The former prescribes macro-level design principles, including \texttt{layout structure} ($L$), \texttt{color system} ($C$), \texttt{shape language} ($S$), and \texttt{usage scenario} ($U$). We define it as a quadruple:

\begin{equation}
\mathcal{G} = \langle L, C, S, U \rangle
\end{equation}

Here, $L$ is a parameterized grid layout description (e.g., $L = Grid(col=12, spacing=8px)$), augmented with semantic organization labels (e.g., ``card-based layout''). $C$ is a set of color vectors paired with semantic roles, such as $(\#CF9BDE, \texttt{Accent})$, ensuring both numerical accuracy and stylistic consistency~\cite{bowen2008formal}. $S$ represents shape language and can be expressed as $(r, f)$, where $r$ denotes a parameterized geometric value (e.g., corner radius $r=12px$) and $f$ denotes a semantic style label (e.g., ``rounded,'' ``minimalist''). Finally, $U$ encodes semantic scenario tags describing interaction rhythms (e.g., ``rapid browsing,'' ``lightweight interaction''), providing contextual anchors for the global specification~\cite{lamine2022understanding}.

Complementary to the global specification, the \textit{Page Composition} captures the recursive decomposition of a page into sections and components, yielding the hierarchical relation:

\begin{equation}
\mathcal{P} = \{\, p_i \mid p_i = \langle Sec_j \rangle \,\}, \quad Sec_j = \{\, comp_k \mid comp_k \in Comp \,\}
\end{equation}

That is, a page $\mathcal{P}$ consists of multiple sections, each containing a set of components, following the structure \texttt{Page} $\rightarrow$ \texttt{Section} $\rightarrow$ \texttt{Component}.  
To avoid ambiguity between instances and schema definitions, we formalize a section schema as:

\begin{equation}
SecSchema = \langle id, pos, layout, color, Comp \rangle
\end{equation}

Here, \texttt{id} provides a unique identifier, allowing direct indexing in a \texttt{React} environment via \texttt{querySelector}. \texttt{pos} specifies relative positioning and proportional relations (e.g., ``left panel 20\%, right main canvas 80\%''). \texttt{layout} encodes parameterized grid descriptions and alignment rules, typically $layout = (grid_{m\times n}, spacing=8px)$, where $m,n$ indicate component arrangement dimensions. \texttt{color} inherits from the global color system $\mathcal{G}.C$, ensuring consistency, while \texttt{Comp} denotes the component set.  

Each component schema can be defined as:

\begin{equation}
CompSchema = \langle type, id, func, layout, color \rangle
\end{equation}

where \texttt{type} belongs to a predefined component set (e.g., \texttt{Menu}, \texttt{Card}, \texttt{Statistic}); \texttt{func} encodes a semantic functional role (e.g., ``navigation,'' ``filter,'' ``summary''); and both \texttt{layout} and \texttt{color} follow the same parameterized–semantic duality as at the section level~\cite{council2001component}.  

Within this framework, a constraint relation is established between the global specification $\mathcal{G}$ and the page composition $\mathcal{P}$:

\begin{equation}
\forall Sec_j \in \mathcal{P}, \quad Sec_j.layout \subseteq \mathcal{G}.L \ \wedge \ Sec_j.color \subseteq \mathcal{G}.C \ \wedge \ Sec_j.shape \subseteq \mathcal{G}.S
\end{equation}

That is, the attributes of each section must be subsets or specializations of the global specification. This inheritance relation enforces global consistency while enabling local flexibility.  

For implementation, we encode SPEC as a lightweight \texttt{JSON} structure: parameterized fields are expressed as numerical or enumerated values (e.g., \texttt{8px}, \texttt{\#FFFFFF}), while semantic fields are stored as short phrases or tags (e.g., ``primary color,'' ``navigation area''). This hybrid design allows large language models to consume deterministic numerical fields for controllable generation, while designers directly manipulate semantic tags to adjust style or intent during iteration~\cite{lamine2022understanding,delgado2016reusing}. Thus, SPEC formalization bridges computational determinism and human-centered interpretability, echoing Bowen and Reeves' call for formal UI models~\cite{bowen2008formal}, aligning with design systems practices summarized by Lamine and Cheng (2022)~\cite{lamine2022understanding}, and remaining consistent with specification-based UI modeling~\cite{heer2010declarative, narechania2020nl4dv}. In practice, we bind this structure to \texttt{React} and \texttt{Ant Design} component frameworks, with \texttt{Recharts} for visualization, thereby creating a verifiable and operational link between generative modeling and front-end engineering.

\subsection{UI SPEC Generation}

\begin{figure}[t]
\centering
\includegraphics[width=1\textwidth]{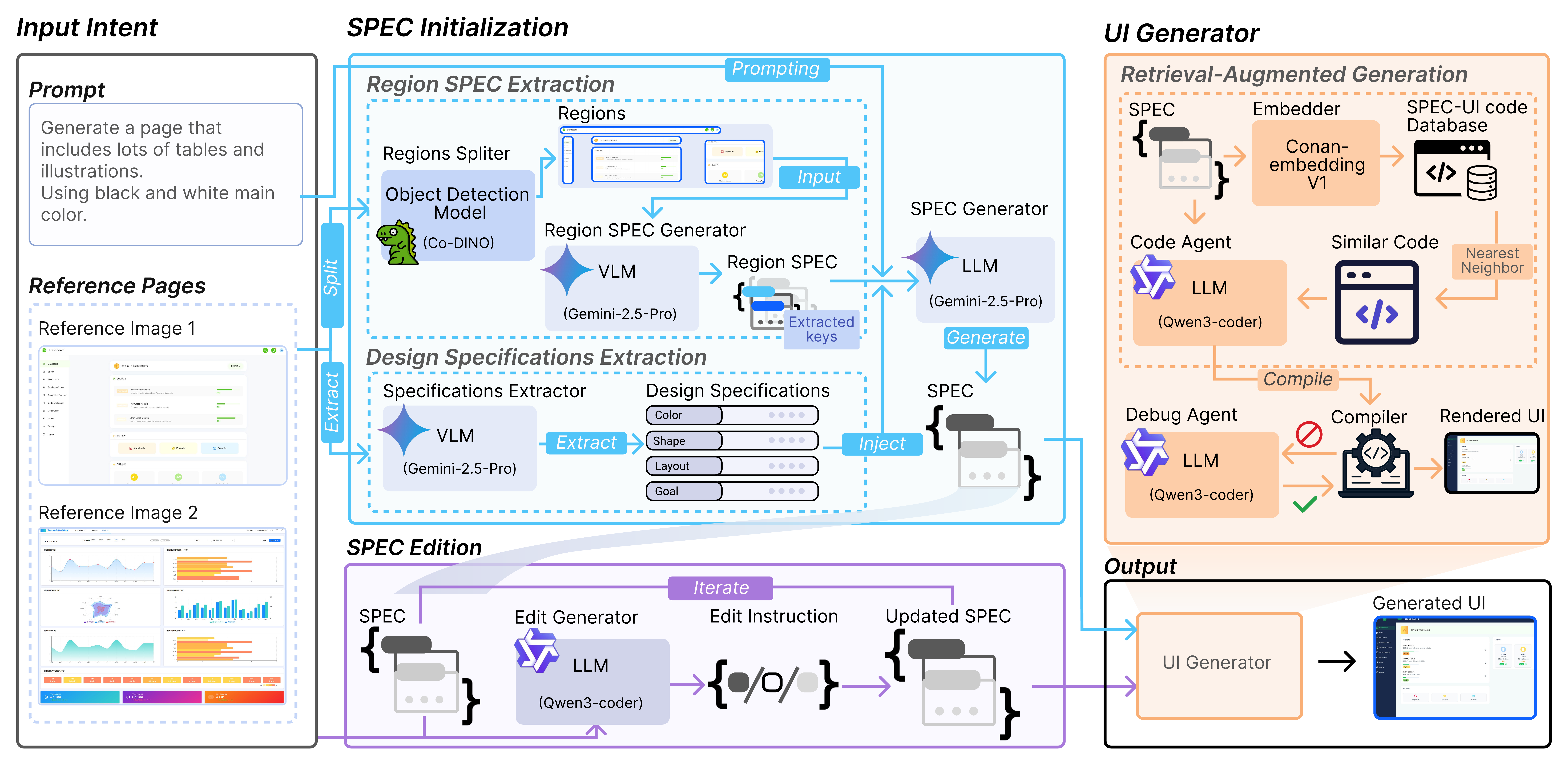}
\vspace{-0.11in}
\caption{\textbf{The technical pipeline of \tool{} transforms user intent into three stages: initializing SPEC, editing SPEC, and finally rendering it into a UI design.}}
\label{fig:Tech_pipeline}
\vspace{-0.1in}
\end{figure}

As shown in Figure \ref{fig:Tech_pipeline}, the generation of SPEC proceeds in three stages: region segmentation, region-wise specification extraction, and global style integration. Together, these stages instantiate the dual-layer structure of SPEC, where the Global Specification defines macro-level design principles, and the Page Composition decomposes the interface into a hierarchy of sections and components. The final representation can be formalized as \textit{S = ⟨G, P⟩}, with \textit{G} capturing global attributes and \textit{P} representing the set of structured region specifications.

\subsubsection{Region Segmentation via Visual Parsing}
To enable structured analysis of reference UIs, we first segment each interface into semantically coherent regions. This preprocessing step transforms a complex layout into manageable visual units, thereby reducing the cognitive and computational load for downstream multimodal LLMs. We train a dedicated detection model, Co-DETR~\cite{zong2023detrs}, from scratch, motivated by its transformer-based collaborative assignment scheme, which is well-suited to the large scale variations in Web UI elements—from full-width banners to small interactive icons.

To support training, we curate a dataset of 8,000 Web UI screenshots and recruit 10 annotators over 7 days to draw bounding boxes over all visible regions. Unlike class-based annotation, we enforce only one rule: the union of all bounding boxes must fully cover the page. This ensures exhaustive coverage while significantly accelerating the annotation process. We choose to train a custom detector rather than rely on existing UI parsers~\cite{wu2025mllm,wu2021screen}, as we observed that off-the-shelf methods often miss content blocks or fragment semantically unified regions, particularly in dense or asymmetrical layouts. Finally, we apply a lightweight post-processing algorithm grounded in Gestalt principles \cite{xie2022psychologically}—merging boxes when alignment, proximity, or enclosure cues suggest visual grouping—while strictly preserving complete page coverage.

% Model training was conducted using the MMDetection framework. We trained for 12 hours on 8×V100 GPUs with a total batch size of 16 (2 samples per GPU). The optimizer used an initial learning rate of 1e-4, and other settings followed standard Co-DETR training protocols.

\subsubsection{Region-wise SPEC Extraction using MLLMs}
After segmentation, each region $r_j \in R$ is passed to a multimodal large language model (MLLM), which generates a structured description that we call a \textit{Region SPEC Unit} (RSU). Conceptually, the model takes as input the cropped region image and a carefully designed prompt, and outputs a JSON-like record of layout, components, functional roles, and stylistic attributes. Formally, we treat each RSU as an instantiation of a \texttt{Section} in the SPEC hierarchy:

\begin{equation}
RSU_j = Sec_j = \langle id_j, pos_j, layout_j, color_j, Comp_j \rangle
\end{equation}

The MLLM’s role is to infer hidden structure from pixels—e.g., recognizing that a rectangular block with text and a button corresponds to a card-like layout—and to translate this into parameterized fields (grid size, spacing, hex colors) and semantic tags (navigation,``primary button'', ``card background''). For instance, when given a product card region, the model may infer a \texttt{grid-1col} layout with one image, a title, a price label, and a button, styled with a white background and an orange accent (e.g., \texttt{\#FF6600}). To increase reliability, we adopt few-shot chain-of-thought prompting~\cite{wei2022chain}, which guides the MLLM to reason step by step: from identifying the high-level section type, to enumerating components, and finally encoding attributes into the RSU format.

\subsubsection{Global Style Extraction and SPEC Integration}
While region-wise analysis captures the internal structure of local areas, many stylistic cues only emerge when the page is considered as a whole. To account for this, we prompt a multimodal LLM with the full UI screenshot to extract a \textit{Global Design Profile}. The model summarizes high-level properties such as overall tone, dominant color palette, and recurring layout rhythms. These attributes provide the global consistency that individual region outputs alone cannot guarantee and are recorded in the \texttt{<global\_specification>} section of the SPEC. Finally, we integrate the compositional SPEC derived from individual regions with the Global Design Profile. The result is a unified SPEC that preserves both the fine-grained decomposition of each region and the stylistic coherence of the entire interface. This consolidated representation serves as the foundation for downstream UI generation, ensuring that local structural details remain consistent with the page’s overall design language. The full prompt is provided in Appendix~A.

\subsection{UI SPEC Edition}

To enable iterative refinement, our system interprets user editing intent—expressed in natural language or via reference images—into structured edit instructions (\autoref{fig:Tech_pipeline}). An LLM acts as the \textit{Edit Generator}, using few-shot prompting to translate user inputs into triplets of the form \texttt{<operation, path, value>}. Here, \texttt{operation} specifies the action (e.g., replace, insert), \texttt{path} identifies the targeted attribute within the SPEC hierarchy, and \texttt{value} encodes the updated content. For example, a user request to adopt a darker theme may be rendered as:
\texttt{<replace, /VisualStyle/DesignStyle, "Tech-oriented dark background with neon-accented charts">}.

Applying such instructions directly modifies only the specified node in the structured SPEC. Because the SPEC is hierarchically organized, this mechanism localizes edits—changing the color palette of a single section, for instance—while preserving the surrounding layout, component composition, and other design attributes. Compared with prompt-based regeneration, which often discards prior structure, this approach provides stable, fine-grained control and ensures consistency across iterations.

To maintain robustness, edits are applied within a controlled execution loop that validates operations, paths, and values. Invalid instructions trigger an exception-handling routine that returns the error context (\texttt{<error message, original instruction, current SPEC>}) to the LLM, prompting it to generate a corrected edit. We cap retries at three attempts to balance responsiveness with stability. This closed-loop mechanism allows the SPEC to evolve safely while retaining its structural integrity, enabling designers to iteratively refine UIs without the risk of global disruptions.

\subsection{UI Page Generation}

The Final SPEC provides a structured blueprint that can be deterministically mapped into executable user interfaces. To achieve this, we implement a UI code generation module (Figure \ref{fig:Tech_pipeline}) composed of three coordinated components: a code agent, a Retrieval-Augmented Generation (RAG) system, and a debug agent. The generator is based on Qwen3-Coder, a state-of-the-art open-source code LLM that offers performance competitive with frontier models (e.g., GPT-4, Claude-4) while enabling fine-tuning and deployment flexibility. We target \texttt{React} as the rendering framework, adopt Ant Design for reusable components, and use ECharts for data visualization. Details of the prompting strategy are provided in Appendix~B.

\subsubsection{Retrieval-Augmented Generation}

Although Qwen3-Coder performs strongly, its outputs can diverge in fidelity and stylistic alignment. To mitigate this, we construct a SPEC–UI Code Database by collecting 2,500 professionally designed web UIs, extracting their SPECs, and generating corresponding React implementations with Claude-4 under controlled prompts. After human validation, 2,000 high-quality SPEC–code pairs are retained. During inference, an input SPEC is embedded and matched against this database; the most similar examples are retrieved and injected into the code prompt as few-shot demonstrations. This grounding process improves both functional alignment and visual coherence while avoiding full reliance on prompt-only learning.

\subsubsection{Debugging and Self-Correction}

Even with SPEC grounding and retrieval support, LLM-generated code may contain syntax errors, missing imports, or hallucinated components. To ensure robustness, we introduce a Debug Agent that executes a compile–feedback–repair loop. After generation, the code is compiled and rendered in a browser runtime; if failures occur, the system captures the error trace and reformulates it into a structured report containing the error type, affected snippet, diagnostic message, and corresponding UI region. This report is appended to the generated code and returned to the LLM for revision. The loop repeats up to three times to balance correction with latency. By incorporating compiler feedback into generation, the Debug Agent ensures that final outputs are both functionally executable and visually faithful to the design intent encoded in the SPEC.

%% file: sections/SystemOverview.tex
\section{\tool: a Generative and Editable Interactive UI Deisgn System}
\subsection{System Overview}
\begin{figure*}[t]
\centering
\includegraphics[width=0.6\textwidth]{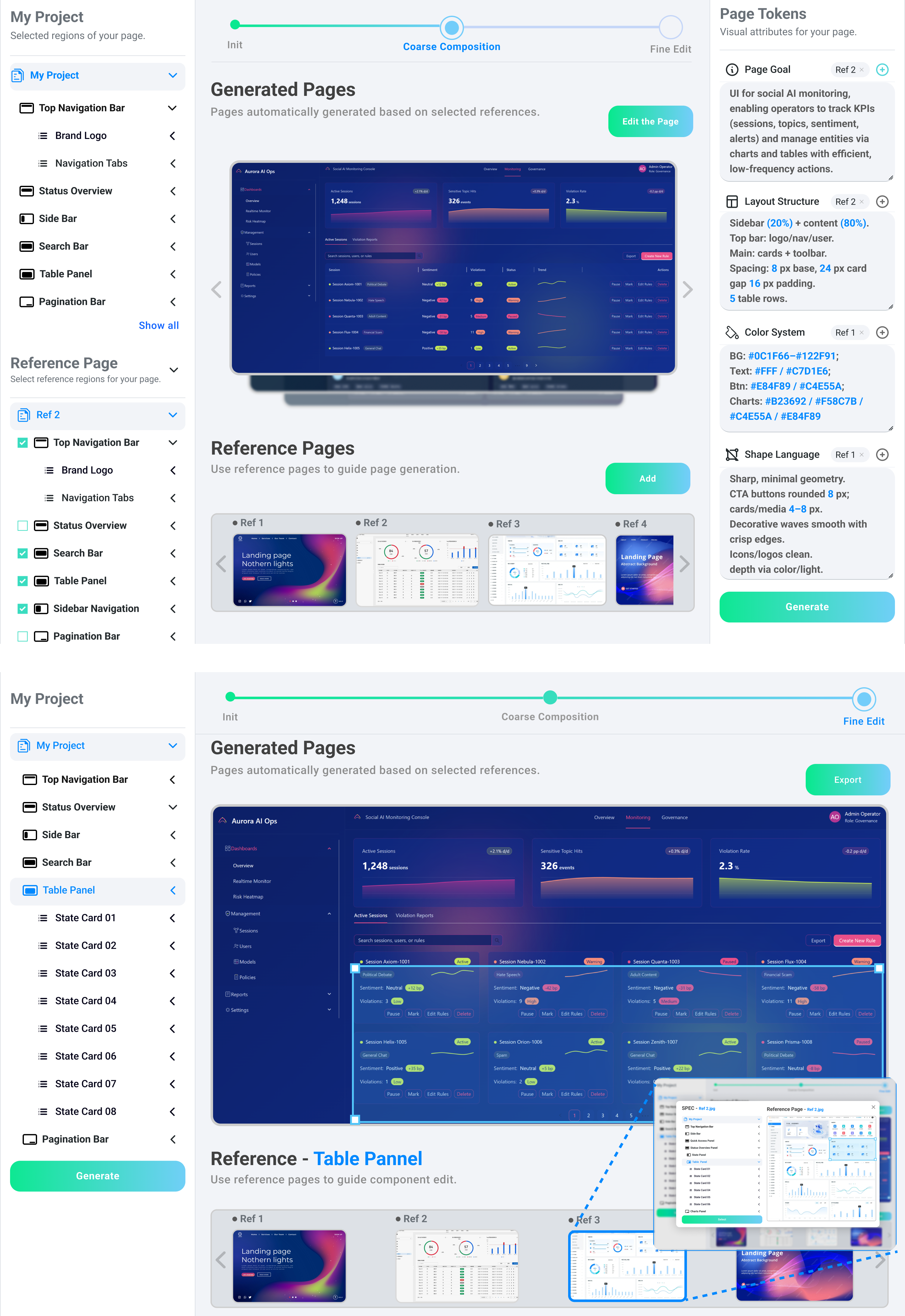}
\vspace{-0.11in}
\caption{\textbf{The user interface of SpecifyUI consists of two panels: (A) the SPEC Composition Panel, which supports extracting design specifications, arranging components, and generating UIs; and (B) the UI Editing Panel, which enables direct manipulation by selecting elements through a visualized UI hierarchy and expressing modification intent with reference images.}}
\label{fig:system}
\vspace{-0.1in}
\end{figure*}

Based on the design goals above, we implemented \textit{SpecifyUI}, an interactive design tool driven by SPEC and a UI generation pipeline. The system is designed to support designers in the early stages of conceptual ideation by combining reference images with their own creative input to generate UI designs. It further enables controllable modifications across iterative cycles, supporting a progression from coarse-grained to fine-grained refinement. Overall, SpecifyUI consists of three primary interface areas: the UI canvas (Fig.~\ref{fig:system}a), the SPEC panel (Fig.~\ref{fig:system}b), and the design specification editor (Fig.~\ref{fig:system}c). During the initial generation phase, designers can upload one or more reference images. The system automatically extracts candidate global specifications and regional structures following the hierarchical schema of SPEC. Designers may then select elements that align with their intended design direction, thereby aligning user intent with LLM-based generation (DG1). Furthermore, SpecifyUI allows users to combine multiple references hierarchically to construct a composite intent representation that guides the initial UI generation (DG2). This process not only encourages exploration of diverse sources of inspiration, but also ensures stylistic and structural coherence through SPEC's structured constraints. For iterative refinement, SpecifyUI exposes the hierarchical structure of the page as defined by SPEC. Designers can directly select elements at the global, regional, or component level, and adjust attributes such as layout, color, or functionality through direct manipulation. The system maintains version histories automatically, enabling designers to quickly roll back or compare design alternatives during exploration (DG3). By grounding editing in SPEC, designers are freed from repeatedly reconstructing complete prompts, and can instead perform efficient, targeted modifications through structured selection and parameterized control.

\subsection{User Scenario}

To demonstrate the use of our system, we present a scenario involving \textit{Sarah}, a junior UI designer. Sarah was tasked with designing a system monitoring page for a social AI platform. Because this type of interface was relatively novel, she struggled to find relevant inspiration and struggled to adapt existing community resources. She decided to use SpecifyUI to bootstrap her design process.

\subsubsection{Reference-based UI Generation.}
Sarah began by gathering three reference images: two network monitoring dashboards (used to inspire layout and component choices; Ref.~A and Ref.~B), and one futuristic technology-themed UI (used to inspire stylistic choices; Ref.~C). She uploaded these references into SpecifyUI, which automatically parsed each image into structured SPEC representations containing both global specifications and hierarchical component structures. When Sarah clicked on a reference, the extracted design tokens appeared in the right-hand panel (\autoref{fig:system}), while the component hierarchy was displayed on the left (\autoref{fig:system}).

To construct her own SPEC, Sarah first filled in the \texttt{PageGoal} to align with her design task. She then reviewed the extracted attributes from Ref.~C: the system highlighted both parameterized values and semantic descriptors, which allowed her to clearly distinguish stylistic properties. Sarah selected the \texttt{ColorSystem} and \texttt{ShapeLanguage} from Ref.~C. Next, she chose the navigation bar and table components from Ref.~A, and the data-monitoring widget from Ref.~B, arranging them into a hierarchy using drag-and-drop. At the global level, she adopted the \texttt{LayoutStructure} from Ref.~A. With this hybrid SPEC assembled, she clicked the \texttt{Generate} button, and SpecifyUI produced an initial UI layout on the canvas. To further refine the design, Sarah directly replaced the background color with a more “tech-inspired” palette, regenerating the interface until she was satisfied.

\subsubsection{SPEC-based Direct Editing.}
Once Sarah had an initial draft, she moved into refinement. By clicking \texttt{Edit Page}, she entered the editing mode. Here, the system exposed the hierarchical SPEC structure in a directory-like view similar to Figma. Each node in the hierarchy was bound to the corresponding visual element, which SpecifyUI highlighted when selected. Sarah wanted to explore two alternatives for the \texttt{Online Sessions} section: one displayed in a table, the other in cards. She selected the component in the hierarchy, opened the attribute panel, and modified the \texttt{Functionality} field by typing “display using cards.” After she clicked \texttt{Edit}, the system regenerated the interface: the table was successfully transformed into a card layout, and the SPEC was automatically updated with the change highlighted.

However, Sarah felt the generated card design was too minimal. To refine it further, she uploaded a new reference image containing an appealing card design. She selected the card element from this reference, and SpecifyUI regenerated the component by integrating its style, functionality, and layout. The new design better matched her intent, and also sparked new creative ideas. Satisfied with the results, Sarah exported the generated UI along with its corresponding code to hand off to engineers for prototyping.

This scenario illustrates how SpecifyUI supports designers in composing hybrid specifications from multiple inspirations, iteratively refining through direct manipulation, and maintaining semantic coherence between intent, structure, and generated outcomes.

%% file: sections/Evaluation.tex
\section{Technical Evaluation}  

The technical evaluation is to test whether our UI generation pipeline can better follow user intent, which in this setting means \textit{reconstructing the reference UI screenshot as faithfully as possible}. We compare against several prompting baselines and then isolate the contributions of three core components in our method: structured specification \textit{(SPEC)}, retrieval augmentation \textit{(RAG)}, and region segmentation \textit{(Region Crop)}. The purpose of this evaluation is not only to measure raw generation quality, but also to verify how each component helps preserve layout, style, and function while keeping the generated UI consistent with the reference design. 

\subsection{Metrics}  
We adopt three complementary metrics to quantify UI generation fidelity. \textbf{Mean Squared Error (MSE)} measures pixel-level reconstruction accuracy, with lower values indicating closer alignment to the target rendering. \textbf{CLIP similarity} \cite{si2025design2code} evaluates semantic alignment between generated and reference UIs in a joint vision–language embedding space, reflecting whether the generated interface conveys the same high-level meaning. \textbf{Structural Similarity Index (SSIM)} captures perceptual and spatial consistency, rewarding preservation of layout and structural patterns beyond raw pixel overlap. Together, these three metrics provide a balanced view of numerical accuracy, semantic faithfulness, and structural coherence.  

\subsection{Baselines}  
We adopt three prompting-based baselines \cite{si2025design2code} to benchmark the effect of structured specification and retrieval grounding. \textbf{Direct Prompt} directly instructs an LLM to generate React code from the UI screenshot without additional preprocessing. \textbf{Text-augmented Prompt} introduces OCR preprocessing, supplying recognized text tokens to the LLM to mitigate content loss, especially for UIs with dense labels. Finally, \textbf{Self-Revision Prompt} extends the text-augmented setup by prompting the LLM to iteratively compare the generated UI with the original screenshot and revise discrepancies, thereby simulating a self-correction loop.

\subsection{SPEC-based Methods}

Beyond prompting baselines, we evaluate our structured generation methods that explicitly use the SPEC representation. These variants allow us to isolate the effect of structure, retrieval, and segmentation:

\textbf{Only SPEC} uses the structured specification alone. This tests whether explicitly encoding layout, hierarchy, and style can already improve fidelity compared to unstructured prompts.

\textbf{SPEC+RAG} augments the structured prompt with retrieval from our SPEC–code database. This examines whether grounding generation with similar design examples improves semantic alignment and style consistency.

\textbf{SPEC+Region Crop} adds region-level cropping to the input, enforcing local spatial constraints. This variant evaluates how spatial decomposition contributes to structural and perceptual fidelity.

\textbf{Integrated SPEC (SPEC+RAG+Region Crop)} combines all three components. This represents our complete pipeline and is used to test whether the three strategies act in a complementary way to maximize intent preservation.

\subsection{Quantitative Results}
Table \ref{tab:quantitative_results} shows that Integrated SPEC achieves the strongest performance across all metrics, reaching an MSE of 40.99, CLIP similarity of 0.887, and SSIM of 0.854. Relative to the strongest baseline (Self-revision Prompt, MSE 50.86 / CLIP 0.755 / SSIM 0.787), this corresponds to an 11.4\% reduction in reconstruction error, and relative gains of +7.6\% in semantic similarity and +14.4\% in structural fidelity. These results establish structured SPEC-guided generation, augmented retrieval, and region-level decomposition as a complementary pipeline for high-fidelity UI synthesis.

Breaking down the contributions of individual components, we observe three key trends. First, structured prompting is the critical driver of improvement. The transition from unstructured prompts to Only SPEC already reduces MSE from 57.42 to 45.14 ($-$21.4\%), while raising CLIP by +22.9\% and SSIM by +18.7\%. This confirms that explicitly encoding layout and hierarchy provides the model with a precise structural scaffold absent in textual baselines.

Second, retrieval augmentation offers further semantic alignment. SPEC+RAG achieves nearly identical structural fidelity to Only SPEC but sustains high CLIP similarity (0.860, +22.5\% over Direct Prompt), validating the role of contextual grounding through design exemplars.

Third, region-level cropping enforces perceptual and spatial coherence. SPEC+Region Crop improves MSE to 42.12 ($-$26.6\%), with consistent gains in CLIP (+24.5\%) and SSIM (+20.6\%), showing that localized decomposition helps the model respect spatial layouts.

Finally, combining retrieval and cropping yields the best of both worlds: \textbf{Integrated SPEC} not only matches the structural accuracy of SPEC+Region Crop but also inherits the semantic grounding of SPEC+RAG, thus producing the highest fidelity across pixel, semantic, and structural dimensions. From a user perspective, this translates to more consistent and faithful UI generation. Designers and developers can expect generated interfaces that reliably preserve intended layout, component hierarchy, and stylistic choices, minimizing the need for manual correction.

\begin{table}[h]
\centering
\caption{
\textbf{Quantitative comparison of different UI generation strategies. Our method (\textbf{Integrated SPEC}) achieves the best results across all metrics, demonstrating improvements in pixel accuracy (MSE), perceptual alignment (CLIP), and structural fidelity (SSIM).}
}
\begin{tabularx}{\textwidth}{l *{3}{>{\centering\arraybackslash}X}}
\toprule
\textbf{Method} & \textbf{MSE}↓ & \textbf{CLIP}↑ & \textbf{SSIM}↑ \\
\midrule
Direct Prompt & 57.4214 & 0.7024 & 0.688 \\
Text-augmented Prompt & 53.2755 & 0.7058 & 0.7017 \\
Self-Revision Prompt & 50.8570 & 0.7553 & 0.787 \\
\rowcolor{gray!10} \textbf{Integrated SPEC (SPEC+RAG+Region Crop)} & \textbf{40.9930} & \textbf{0.8871} & \textbf{0.854} \\
SPEC+RAG & 44.8892 & 0.8598 & 0.8167 \\
SPEC+Region Crop & 42.1189 & 0.8744 & 0.8302 \\
Only SPEC & 45.1431 & 0.8628 & 0.8166 \\
\bottomrule
\end{tabularx}
\label{tab:quantitative_results}
\end{table}

\subsection{Qualitative Results}

\begin{figure*}[t]
\centering
\includegraphics[width=0.9\textwidth]{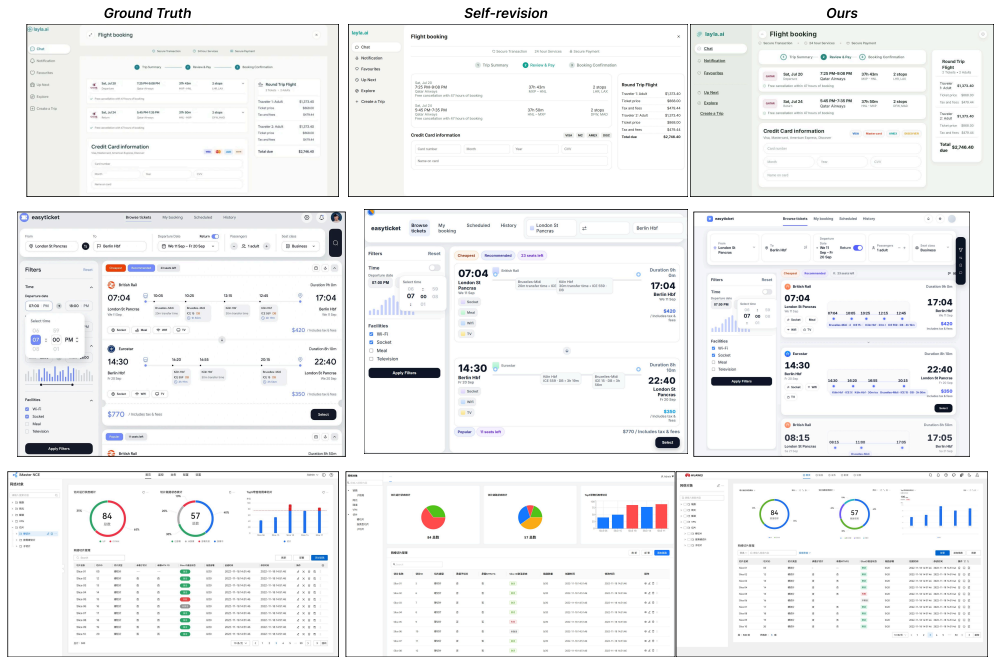}
\vspace{-0.15in}
\caption{\textbf{Qualitative results of the UI generation pipeline compared with the best baseline method for fidelity-oriented design generation.}}
\label{fig:Quali}
\vspace{-0.1in}
\end{figure*}

\section{User Study}

Building on the technical evidence of intent alignment, the user study evaluates whether these gains translate into human-perceived benefits—output quality (RQ1), intent expression and controllability (RQ2), and interaction experience and effort (RQ3).:

\begin{itemize}
    \item \textbf{RQ1: Design Output Quality.}  
    How effectively does SpecifyUI support UI designers in producing high-quality UI designs compared to Google Stitch as a commercial baseline?

    \item \textbf{RQ2: Intent Expression and Control.}  
    In what ways does SpecifyUI support designers’ intent expression and controllability during both UI generation and iterative editing, compared to Google Stitch?

    \item \textbf{RQ3: Interaction Experience.}  
    How do designers’ experiences and behaviors differ between SpecifyUI and Google Stitch?
\end{itemize}

\subsection{Method}
\subsubsection{Participants}
We recruited a total of 16 participants (9 female, 7 male), consisting of both university students and professional designers from industry. Participants were invited through word-of-mouth, and all were active users of generative AI tools (e.g., ChatGPT) in their daily work or study. This sampling strategy ensured that participants had prior familiarity with generative interaction while also representing designer communities across both academic and industrial contexts.

\subsubsection{Basline}
We compared SpecifyUI against the industry-standard tool Google Stitch, which we used as a benchmark. Google Stitch is a state-of-the-art UI generation system powered by the Gemini 2.5-pro large model. It allows users to generate high-quality UIs and corresponding code from either text prompts or image prompts. The interface consists of a chat panel on the left and a flexible canvas on the right for displaying the generated UIs.

We also considered other tools as potential baselines, including Claude Artifact and Uizard. However, we found that Claude Artifact was less reliable than Google Stitch in generating UI designs, while Uizard functioned more as an AI-assisted UI design tool rather than a direct UI generation system. Ultimately, we identified Google Stitch as the most suitable benchmark for our evaluation. It supports reference-based UI generation, iterative editing on selected UI elements, and visual tracking of each design version on the canvas. This enables designers to generate multiple alternatives during the creative process, select the most promising one, and refine it through iterative improvements—mirroring the natural way people interact with LLMs and providing an intuitive pathway to explore new possibilities within the design space.

\subsubsection{Study Procedure}
The study lasted approximately 80 minutes. 7 sessions were conducted face-to-face in the laboratory, while 9 were conducted remotely via Tencent Meeting. At the beginning of each session, the experimenter provided oral instructions and ensured that participants signed an informed consent form. The entire process was screen- and audio-recorded, and each participant received \$20 as compensation. We adopted a within-subjects design: each participant used both systems (SpecifyUI and Google Stitch) to complete two design tasks. The order of systems and tasks was counterbalanced to minimize learning effects.

We designed a two-step reference-based task to ensure participants could experience both the generative and multi-granularity editing capabilities of SpecifyUI, while consistently grounding their design intentions in explicit reference UIs. \textbf{Step 1: Reference-based Generation.} Participants first selected several reference UIs, identified key elements (e.g., overall layout, style, or content structure), and used the system to generate an initial interface. This step examined whether the system could effectively transform abstract reference intentions into executable UIs while preserving emphasized features. \textbf{Step 2: Reference-based Editing at Multiple Granularities.} Participants then edited the generated results based on the chosen references, aiming to make the design closer to the selected UIs. Editing was performed at three levels: the global level (e.g., adjusting overall page layout), the regional level (e.g., modifying headers or sidebars), and the component level (e.g., adjusting fonts, button styles, or spacing).

The two design tasks were derived from real-world HCI scenarios discussed at CHI and UIST. \textbf{Task I} required designing a data monitoring dashboard for a social AI system \cite{wang2025social}. \textbf{Task II} asked participants to design a safe, engaging, and easy-to-use travel service webpage for children and their parents \cite{chen2025sceniclocationbasedfostercognitive}. For each task, participants were provided with 10 reference UIs. After a short 10-minute training session, they had 20 minutes to complete the design ideation. At the end of each task, participants selected the UI they found most satisfactory and inspiring as their final submission. After completing both tasks, participants filled out a questionnaire and participated in a semi-structured interview, providing feedback on both systems in terms of creativity support, sense of control, and overall user experience.

\begin{figure*}[t]
\centering
\includegraphics[width=0.9\textwidth]{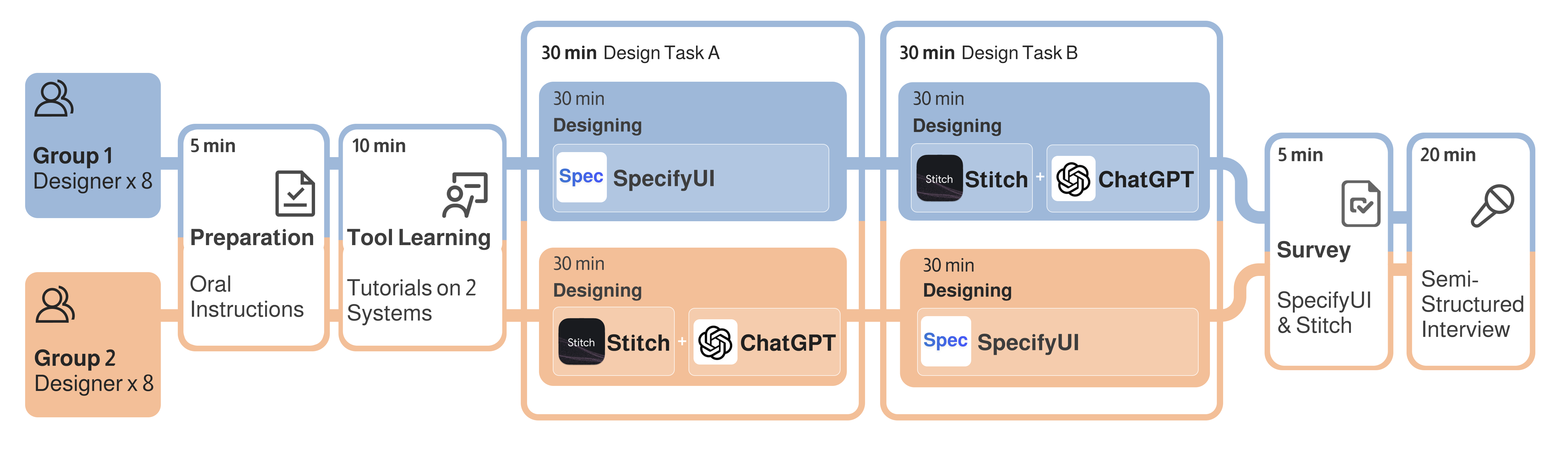}
\vspace{-0.11in}
\caption{\textbf{User Study Procedure}}
\label{fig:User Study Procedure}
\vspace{-0.1in}
\end{figure*}

\subsubsection{Measures}  
To evaluate the effectiveness of \tool{} in relation to our research questions, we employed a combination of expert assessments, participant ratings, and objective interaction logs. Specifically, RQ1 focused on expert-assessed output quality, RQ2 on participants’ perceived ability to express intent and maintain control during design tasks, and RQ3 on broader interaction experiences shaped by different paradigms.  

\textbf{RQ1 (Design Output Quality).}  
We invited three expert UI designers, independent from the study participants, to evaluate the generated UIs. In total, 32 design samples were assessed, covering outputs from both \tool{} and Stitch, with task order counterbalanced across participants. Each design was evaluated along four reference-related dimensions: (i) \textit{style consistency}, (ii) \textit{layout consistency}, (iii) \textit{component correctness}, and (iv) \textit{task relevance}. Experts rated each dimension on a 1–7 Likert scale (1 = strongly inconsistent, 7 = highly consistent). To minimize bias, outputs were anonymized and randomly ordered before evaluation. The ratings from the three experts were averaged for analysis. Inter-rater reliability was also calculated, yielding a Cohen’s $\kappa \approx 0.72$, which indicates acceptable agreement among experts.  

\textbf{RQ2 (Intent Expression and Control).}  
To examine how \tool{} supports designers in expressing intent and maintaining task-level controllability compared to Google Stitch, we adapted evaluation items from established constructs on controllability and perceived ease of use \cite{lallemand2015user, davis1989technology}. Participants rated five aspects on a 7-point Likert scale (1 = strongly disagree, 7 = strongly agree): (i) \textit{controllability in generation}—whether they could effectively steer the system to produce outputs aligned with reference UIs; (ii) \textit{controllability in editing}—whether they could flexibly refine results at global, region, and component levels without unintended changes; (iii) \textit{intent adherence}—whether the generated UIs faithfully followed the design intent conveyed by reference images; (iv) \textit{articulation efficiency}—whether design intentions could be expressed quickly and translated into satisfactory results; and (v) \textit{ease of intent communication}—whether participants felt it was smooth and convenient to convey references or high-level ideas to the system. In addition to subjective ratings, we logged task performance indicators, including time to complete the initial generation and refinement.

\textbf{RQ3 (Interaction Experience).}  
To compare SpecifyUI’s vision-centered interaction paradigm with Stitch’s text-centered paradigm, we examined participants’ subjective experience and objective interaction behaviors. Subjective experience was measured through six items on a 7-point Likert scale (1 = strongly disagree, 7 = strongly agree): (i) overall ease of use, (ii) perceived creative control—whether the interaction modality allowed designers to remain in the lead rather than being directed by the system, (iii) workflow alignment—whether the interaction style matched everyday design practices, (iv) willingness to adopt—whether participants would continue using the system in future real design tasks, and (v) task work load. These items were adapted from established usability and technology acceptance constructs \cite{davis1989technology}. For objective measures, we logged typing effort, quantified as the total number of characters typed. To better capture differences in interaction patterns, typing effort was further categorized by edit scope (global, region, component) and edit type (style, layout, functionality). All subjective evaluations were made on 7-point Likert scales. Since each participant and each design task produced paired comparisons between SpecifyUI and Stitch, we employed paired two-sample t-tests for analysis. For robustness, we additionally ran Wilcoxon signed-rank tests on Likert-scale data, which yielded consistent results.

\begin{figure*}[t]
\centering
\includegraphics[width=0.9\textwidth]{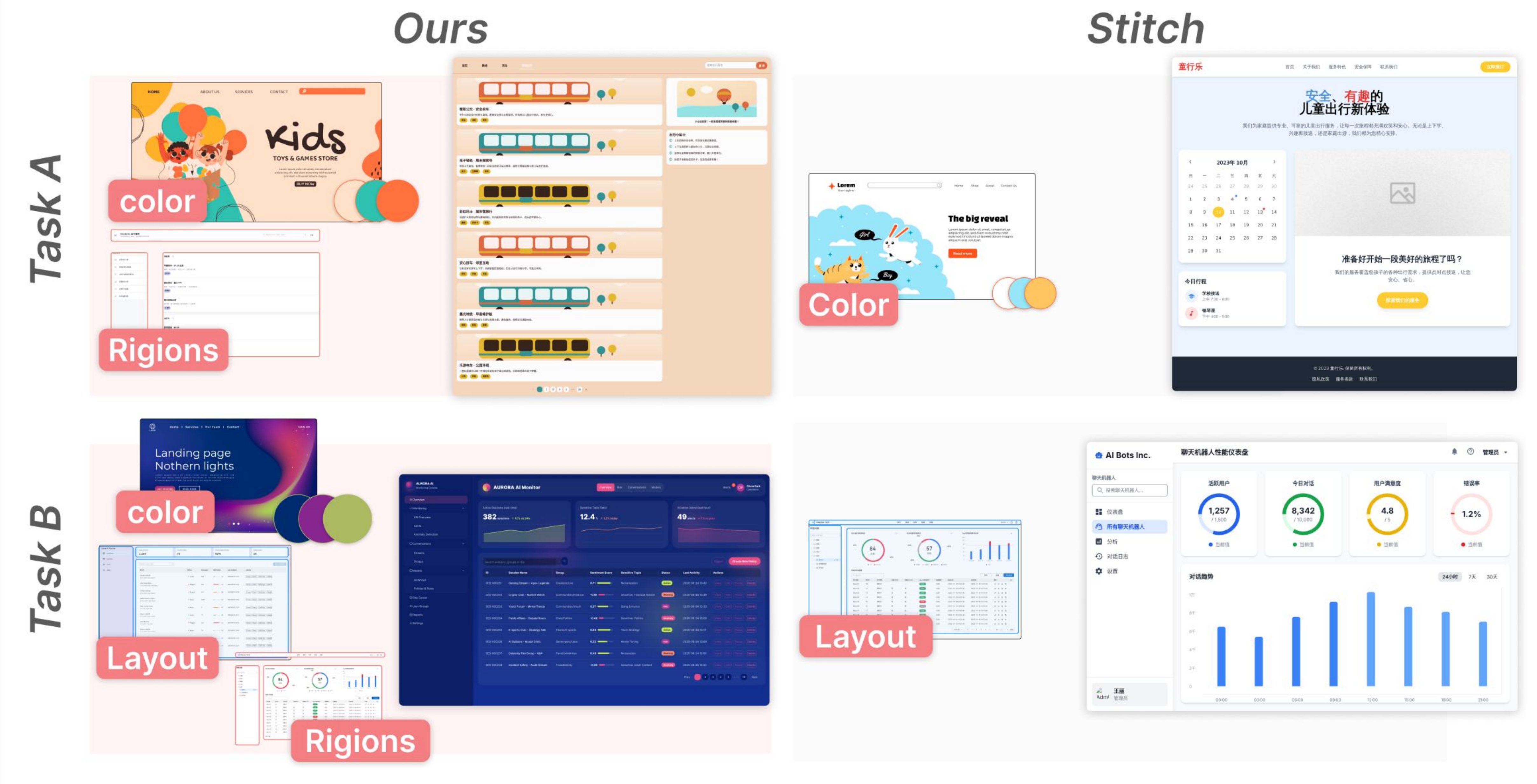}
\vspace{-0.15in}
\caption{\textbf{Examples of participant design processes using \textit{SpecifyUI} and \textit{Stitch}, illustrating user inputs of color, layout, and region component references.}}
\label{fig:user_example}
\vspace{-0.1in}
\end{figure*}

\subsection{Results}
\subsubsection{RQ1: Design Output Quality.}

\begin{figure}
    \centering
    \includegraphics[width=1\linewidth]{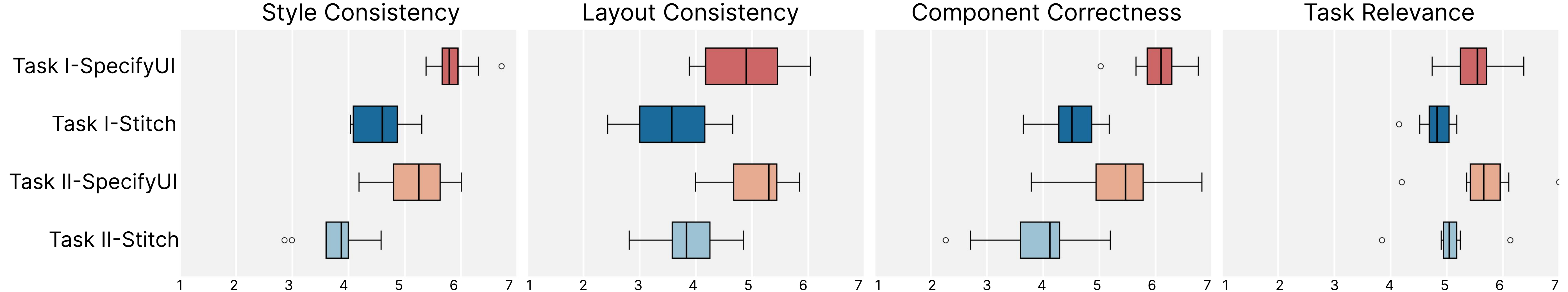}
    \caption{\textbf{Box plots of expert ratings on style consistency, layout consistency, component correctness, and task relevance, showing consistently higher scores for \tool{} over Stitch.}}
    \label{fig:RQ1}
\end{figure}

Figure~\ref{fig:RQ1} shows box plots that clearly indicate higher medians and more compact distributions for \tool{} across all four dimensions. We conducted a paired evaluation of 32 design samples (16 produced with \tool{} and 16 with Stitch), and averaged the ratings of three independent expert UI designers. Each design was rated along four reference-related dimensions: \textit{style consistency}, \textit{layout consistency}, \textit{component correctness}, and \textit{task relevance}. Outputs were anonymized and randomly ordered to minimize bias. Paired $t$-tests revealed that \tool{} significantly outperformed Stitch on all four dimensions (\textit{all $p<.001$}). Specifically, style consistency ratings were higher for \tool{} ($M=5.52$, $SD=0.64$) than for Stitch ($M=4.18$, $SD=0.68$), mean difference $=1.34$, $t(7)=14.79$, $p=1.55\times10^{-6}$; layout consistency also favored \tool{} ($M=4.99$, $SD=0.72$ vs.\ $M=3.73$, $SD=0.71$), mean difference $=1.26$, $t(7)=15.76$, $p=1.00\times10^{-6}$; component correctness showed the largest gap ($M=5.69$, $SD=0.90$ vs.\ $M=4.20$, $SD=0.81$), mean difference $=1.49$, $t(7)=21.99$, $p=1.01\times10^{-7}$; and task relevance was also significantly higher for \tool{} ($M=5.58$, $SD=0.61$ vs.\ $M=4.92$, $SD=0.49$), mean difference $=0.66$, $t(7)=7.42$, $p=1.47\times10^{-4}$. Together, these results indicate that \tool{} not only achieves statistically significant improvements but also yields practically meaningful gains, with the strongest advantages in style fidelity and component correctness.

\subsubsection{RQ2: Intent Expression and Control.}
\begin{figure}
    \centering
    \includegraphics[width=0.8\linewidth]{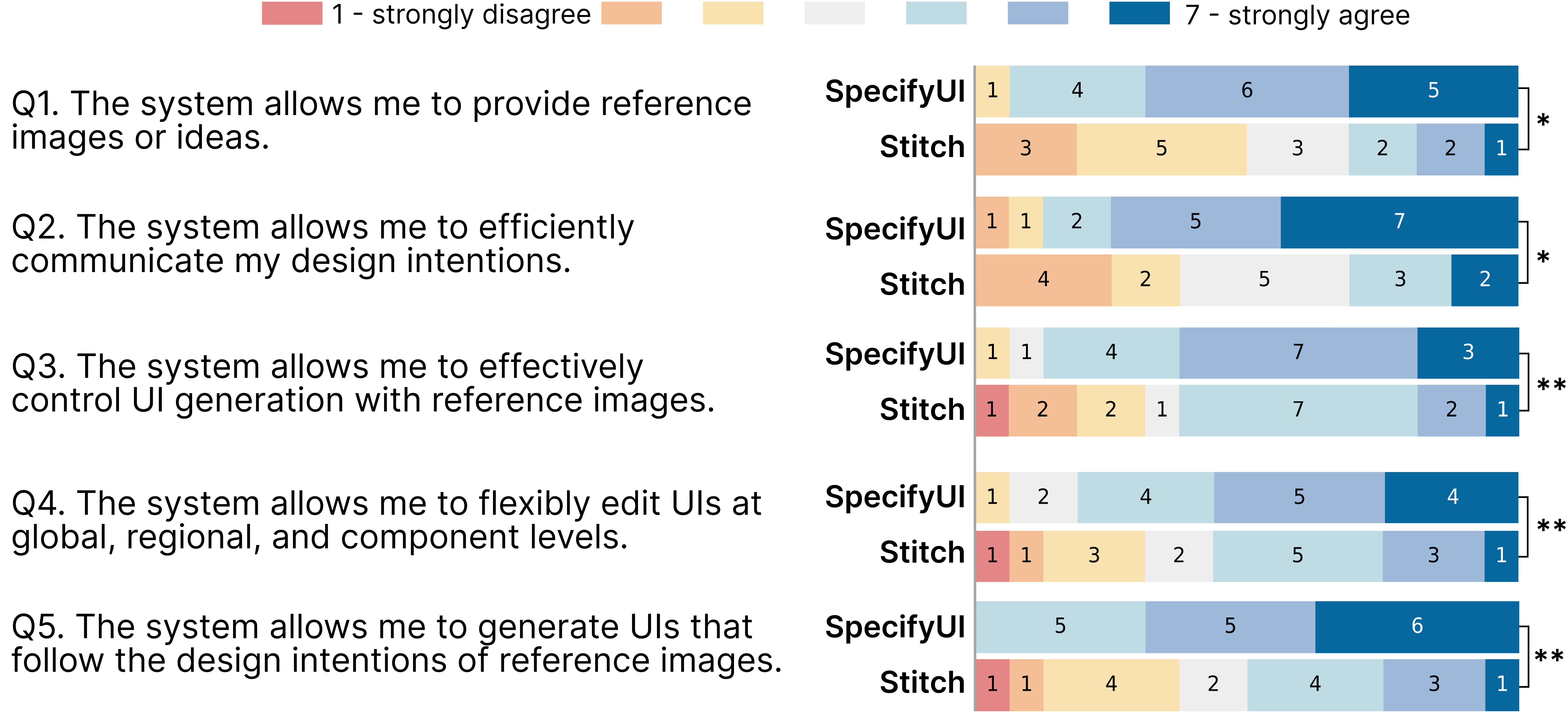}
    \caption{\textbf{Subjective ratings for intent expression and control, comparing \tool{} and Stitch across five dimensions: controllability in generation, controllability in editing, intent adherence, articulation efficiency, and ease of intent communication.}}
    \label{fig:RQ2}
\end{figure}

\begin{figure}
    \centering
    \includegraphics[width=1\linewidth]{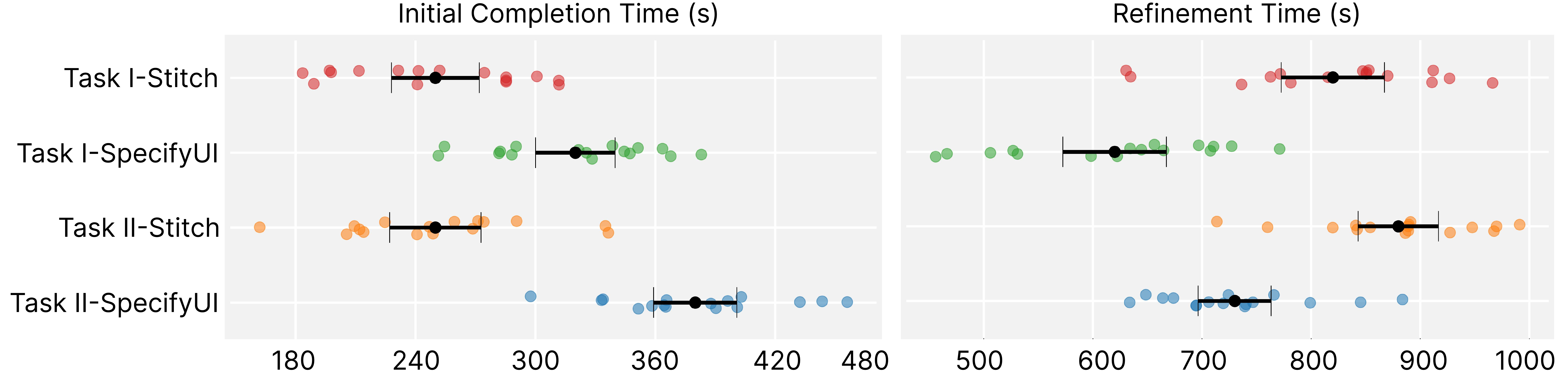}
    \caption{\textbf{Task completion times for initial generation and iterative refinement across two systems.}}
    \label{fig:usage_time}
\end{figure}

% RQ2分析      
Figure~\ref{fig:RQ2} and Figure~\ref{fig:usage_time} summarize the results comparing \tool{} with Google Stitch on intent expression, controllability, and task performance. Overall, participants consistently rated \tool{} higher across all five subjective dimensions, and task performance analysis revealed a trade-off between longer initial setup and faster refinement. More specifically, participants reported stronger controllability during the generation phase with \tool{} ($M=5.88$, $SD=1.05$) compared to Stitch ($M=3.88$, $SD=1.25$; $t=-3.46$, $p=0.0035$), and greater flexibility in editing at global, region, and component levels ($M=5.88$, $SD=1.02$ vs. $M=3.94$, $SD=1.19$; $t=-3.18$, $p=0.0062$). They also rated \tool{} outputs as more faithfully adhering to reference design intent ($M=6.06$, $SD=0.93$ vs. $M=4.25$, $SD=1.29$; $t=-3.88$, $p=0.0015$). In addition, \tool{} was considered more efficient in articulating design intentions ($M=5.56$, $SD=1.15$ vs. $M=4.38$, $SD=1.09$; $t=-2.45$, $p=0.0271$) and smoother in conveying references or high-level ideas ($M=5.63$, $SD=1.07$ vs. $M=4.31$, $SD=1.21$; $t=-2.51$, $p=0.0239$). Non-parametric tests confirmed these findings (all $p<.05$). For task performance (Figure~\ref{fig:usage_time}), \tool{} required more time during the initial generation phase (Task1: $M=380$s vs. $M=250$s, $t=8.24$, $p<.001$; Task2: $M=320$s vs. $M=250$s, $t=4.63$, $p<.001$) due to explicit SPEC construction. However, this upfront effort was compensated by significantly faster refinement (Task1: $M=730$s vs. $M=880$s, $t=-5.91$, $p<.001$; Task2: $M=620$s vs. $M=820$s, $t=-5.84$, $p<.001$). As a result, the total task time did not differ significantly, suggesting that structured intent modeling ultimately streamlined the iterative workflow.  

\subsubsection{RQ3: Interaction Experience.} 

\begin{figure*}
    \centering
    \includegraphics[width=0.8\linewidth]{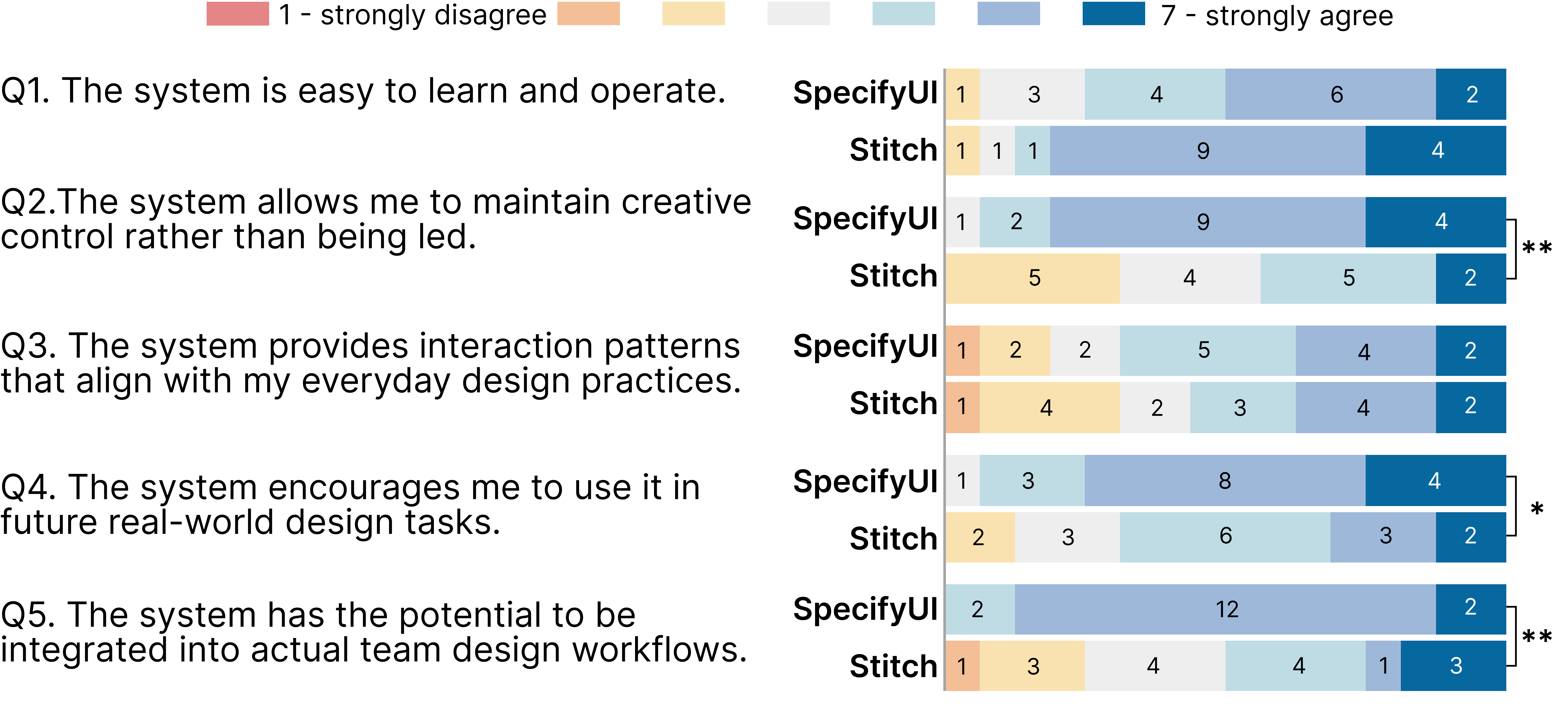}
    \caption{\textbf{Subjective ratings of interaction experience comparing \tool{} and Stitch across five dimensions.}}
    \label{fig:RQ3}
\end{figure*}

\begin{figure}
    \centering
    \includegraphics[width=0.55\linewidth]{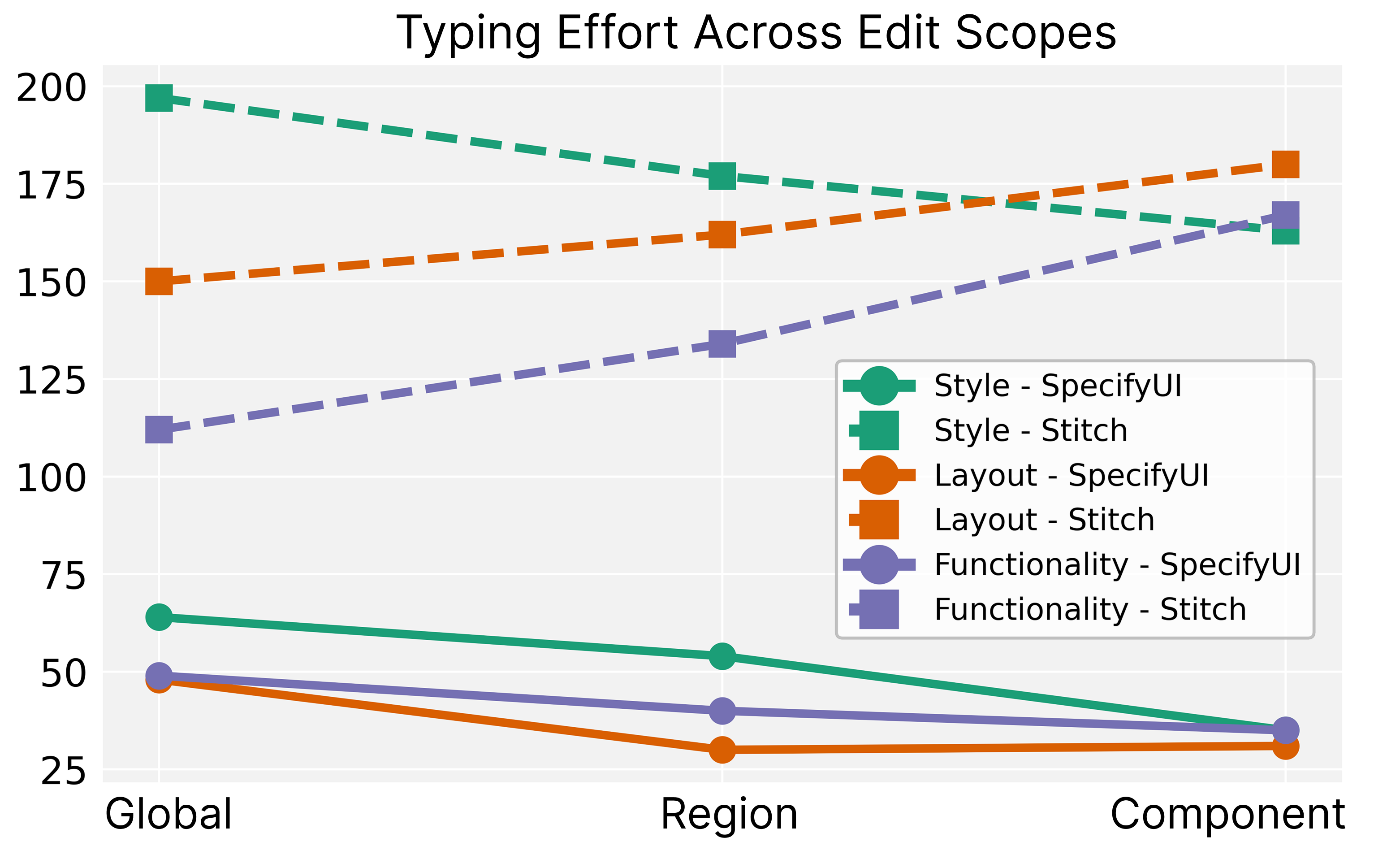}
    \caption{\textbf{Objective typing effort across edit scopes (global, region, component) and content types (style, layout, functionality).}}
    \label{fig:typing}
\end{figure}

Figure~\ref{fig:RQ3} summarizes the subjective ratings comparing \tool{}’s vision-centered interaction paradigm with Stitch’s text-centered paradigm. Participants reported that \tool{} provided significantly stronger creative control ($M=6.00$ vs. $M=4.38$, $t=4.62$, $p<.001$), higher willingness to adopt for future tasks ($M=5.94$ vs. $M=5.00$, $t=2.70$, $p=.017$), and greater potential for integration into team workflows ($M=6.00$ vs. $M=4.63$, $t=3.56$, $p=.003$). No significant differences were observed in overall ease of use ($M=5.31$ vs. $M=5.88$, $t=-1.38$, $p=.19$), likely due to participants’ prior familiarity with text-based prompting. For workflow alignment ($M=4.94$ vs. $M=4.69$, $t=0.50$, $p=.63$), both systems were considered capable of producing usable designs, but \tool{}’s slightly higher score suggests stronger alignment with professional design practices. Objective interaction logs (Figure~\ref{fig:typing}) revealed a clear efficiency advantage for \tool{}. On average, participants typed 386 characters when using \tool{}, compared to 1442 characters with Stitch—an overall reduction of approximately 73\%. This reduction was consistent across all edit scopes and content types, with the strongest savings observed in component-level edits (about 80\% less input) and layout modifications (about 78\% less input). These results suggest that \tool{} substantially lowers interaction cost and better supports fine-grained structural refinement than text-centered prompting.

\subsection{Participants feedback}
In this section, we discuss participants’ feedback on their experience using \tool{} and the Stitch.  

\textbf{Decomposing references into design specifications.} 
Participants found that breaking down reference UIs into multiple dimensions and hierarchical structures was more effective than relying on whole-image prompts, as it better aligned with their professional workflows. As P2 explained, \textit{``Being able to break a UI into parts lets me freely recombine them, instead of being restricted by the original composition.''} Compared to Stitch, which only accepts whole images, \tool{} allowed participants to reuse elements across references (e.g., style from one design and layout from another), making intent expression more precise and flexible. Several participants emphasized that this process mirrored how they normally separate concerns in tools like Figma, where style, layout, and components are independently adjusted.  

\textbf{Articulating design intent.} 
Many participants reported that the automatic extraction of design features from reference images helped them articulate intent more accurately and reduced the burden of verbal description. As P12 noted, \textit{``I know what style or layout I want from a reference, but I don’t know how to describe it. Seeing the extracted design features gave me exactly the words I needed.''} This capability was especially valuable for participants who struggled with technical terminology. As shown in Figure~\ref{fig:RQ2}, the resulting outputs from \tool{} were perceived as more faithful to reference intent, demonstrating improved alignment between high-level ideas and generated results. Some acknowledged that selecting and validating features took more steps, but described the effort as worthwhile because it enhanced precision and reduced ambiguity.  

\textbf{Enhanced controllability but requiring clear goals.}
Participants consistently described \tool{} as providing stronger controllability compared to natural language prompting, especially for editing at global, region, and component levels. As P4 put it, \textit{``With \tool{}, I can make very targeted changes—just the layout or just a component—without messing up the rest of the design.''} This controllability was tied to predictability: unlike Stitch, where prompt tweaks could lead to large and unexpected changes, edits in \tool{} stayed scoped and consistent. However, participants also noted that this advantage presupposed having a relatively clear design goal. When exploring early-stage ideas, Stitch’s natural language interface was perceived as more flexible and faster to spark inspiration. As P11 explained, \textit{``When I don’t know what I want yet, just typing something random in Stitch often gives me new ideas.''}  

\textbf{Leveraging AI for inspiration.}
Across both tasks, participants saw AI not as a tool for final delivery but as a collaborator for generating design directions. They highlighted that \tool{}’s layered specification (Global–Region–Component) provided a shared language for working with the AI, making the collaboration more structured. As P5 summarized, \textit{``Sometimes I just want to try out multiple layouts or data visualizations quickly. In Stitch, I need to re-prompt over and over, but in \tool{} I can just tweak the spec and get variations that make sense.''} Designers with different priorities—whether focused on style, layout, or functionality—felt that \tool{} better supported combining references with explicit design goals, producing outputs closer to production-ready designs.  

% 
% 将参考图分解为设计规范和层级结构比整体使用更有效：参与者一致认为，将参考UI分解为不同维度而非整体使用，更符合他们的设计流程。P2指出：能够把UI拆解成部分，让我能够自由组合，从而契合我的构思，而不是受限于原始构图。与Stitch相比（只能上传整体图像进行参考），参与者发现能够重新组合多个设计维度的参考并明确不同粒度（全局、区域、组件）的参考，对表达设计意图更有效。
% 设计意图表达：多数参与者发现，通过选择参考图并自动提取设计特征，他们能够更准确地表达设计意图。P12提到：我能在参考图中找到自己想要的风格和布局，但我不知道该如何描述它的设计，看到提取的文本化设计特征后，我觉得这就是我需要的。如图x所示，使用SpeciUI，参与者最终产出的UI设计会更契合参考图，这表明系统提高了意图与AI生成结果的一致性。一些参与者指出，逐一提取、审查与选择设计特征的过程虽然复杂且耗时，但他们认为这种付出值得的，因为它能够带来更高的可控性。这也是SpecifyUI在易用性与工作负担未能展现显著优势的原因。
% SpecifyUI增强控制感，但是需要明确目标：参与者一致认为，与自然语言提示相比，SpecifyUI提供了更强的控制感，尤其适合定义UI布局、风格以及特定的编辑。P4评论道：xxx。然而，参与者指出，这种增强的控制感前提是用户已有明确的设计或编辑意图，当他们对想要设计的内容尚不清晰时，Stitch所采用的自然语言提示更灵活，快捷，能够快速进行灵感探索。P11解释道：xxx。这些反馈解释了控制与探索之间的权衡，自然语言提示有助于广泛探索，能从AI反馈中获得灵感，因为AI具有随机性。而SpecifyUI在用户具有大致方向和明确的设计目标时，则在一致性与编辑性上更具优势，正如P6所言：xxx。
% 
% 促进与AI的协作方式：在方案初期，用户希望AI能生成多个UI设计案例，AI的价值在于提供方向性参考而非最终交付。设计师往往关注点不同，有的关注组件功能和布局，有的更关注色彩，这取决于他们的设计需求。P5指出：有时候我想快速尝试多种布局或者不同的数据可视化方式。在Stitch中，这种需求往往导致用户大量反复输入。但在SpecifyUI中，用户可以更轻松地将参考和自己的设计需求结合，从而提供可落地的结果。
% 

%% file: sections/Discussion.tex
\section{Disccussion}
\subsection{Structured Specifications as a Foundation for Human–AI Co-Creation}
% 这一节聚焦在 SPEC 的核心价值：作为中间表示，解决 LLM 驱动 UI 生成中的“模糊表达”问题，强调 intent clarity 和 control。
% 论述围绕两个关键模糊点（意图描述、参考利用），最后上升到 “动态能动性分配” 和 “结构化协作协议”。
SPEC addresses two critical ambiguities in LLM-centered UI generation: first, how to clearly describe a page so that UI design intent can be accurately conveyed to the LLM; and second, how to construct such intent from reference resources to achieve controllable generation and editing. Prior work, such as DirectGPT \cite{DirectGPT} and Brickify \cite{Brickify} attempted to enrich visual exploration by embedding visual symbols or supporting spatial manipulations. However, these purely visual representations remain ambiguous when communicating intent to LLMs, particularly in capturing hierarchical relationships and overlapping layouts. The strength of SPEC lies in mapping these ambiguous visual elements into a hierarchical, editable, and structured specification, allowing design intent to be expressed and operationalized with greater clarity.

From a broader perspective, SPEC is not merely an engineering solution for UI design but also offers a general paradigm for generative interaction systems: introducing a structured intermediate semantic layer between the ambiguity of natural language interaction and the rigidity of code implementation, serving as a “shared language” for human–AI collaboration. This layer captures designers’ high-level semantic intent while simultaneously translating it into executable generation and editing operations, thus establishing a continuous transition between exploratory ambiguity and precise control. For the HCI community, this paradigm highlights two insights. First, dynamic agency allocation: agency flexibly shifts between user and AI depending on the design stage—AI supports broad exploration early on, while SPEC enables precise user control during refinement. This resonates with prior discussions on the dynamic nature of agency (e.g., Satyanarayan et al. \cite{satyanarayan2024intelligence}). Second, structured collaboration protocols: as an intermediate representation, SPEC anchors both human–AI interaction and multi-agent processes (e.g., UI understanding \cite{wang2023enabling}, retrieval \cite{park2025leveraging}, code generation \cite{gui2025latcoder}), making AI a transparent co-creator rather than a black box.

\subsection{Beyond Prompts: Natural Interactions for Conveying Design Intent}
In SpecifyUI, SPEC currently serves as the shared medium between users and the AI, with reference images and textual prompts as the primary channels for intent expression. However, our findings suggest that these modalities are not always sufficient. In early ideation, participants often preferred to begin with natural language prompts when they lacked a concrete vision, aligning with prior observations that text can facilitate divergent exploration \cite{chen2024autospark, wadinambiarachchi2024effects}. In later refinement stages, however, they found natural language edits cumbersome and ambiguous, instead favoring direct visual interactions such as region selection or component manipulation, consistent with evidence that visual modalities better support precision and controllability \cite{suh2024luminate, DirectGPT}. While reference images remain a common entry point, designers also rely on memory or externalize their ideas through quick sketches, which serve as a natural way to articulate intent \cite{jonson2005design, kirsh2010thinking}. These observations, echoed in our study where ease of intent communication received only moderate scores, highlight the need for more natural input channels that allow designers to fluidly shift between textual, visual, and structural forms of expression depending on their goals. Sketch-based modalities are especially promising: rough layout sketches can convey spatial organization, stylistic sketches with color blocks or shape strokes can communicate mood and theme, and interaction sketches using arrows or annotations can indicate intended functionality. Yet the challenge of translating such non-verbal inputs into structured representations for generative models remains underexplored, particularly for local refinements during detailed design. Extending SpecifyUI to incorporate sketch-based and other non-verbal modalities—such as node dragging \cite{lu2025misty}, paint-based area selection \cite{chung2022talebrush}, or line sketching—represents a compelling direction for supporting more natural, flexible, and expressive design workflows.

\subsection{From Design to Prototyping: Bridging with Engineering and Domain Knowledge}
A persistent challenge in generative UI research is the disconnect between AI-generated designs and production-ready implementations \cite{GenUIStudy}. Current UI generation systems largely emphasize design ideation and creative exploration, but they often neglect the critical transition from design concepts to usable, executable prototypes \cite{jing2023layout, gupta2021layouttransformer}. This “last-mile gap” limits the adoption of generative tools in real-world workflows. SpecifyUI addresses this gap not only through SPEC as an intermediate representation but also through its multi-agent generation pipeline. The pipeline coordinates UI understanding, knowledge retrieval, code generation, and self-debugging, ensuring that design intent captured in SPEC can be consistently translated into executable prototypes. Instead of producing static visuals, the system yields editable code artifacts, reducing the manual overhead for engineers and allowing teams to focus on higher-level interaction logic and functionality. Participants in our study emphasized that this continuity reduced the burden of “pixel-perfect handoff” and made the results feel closer to deployable prototypes.

Beyond improving fidelity, specification-driven pipelines open opportunities to integrate external knowledge and reusable resources. By linking SPEC with established component libraries (e.g., Material Design, Ant Design) or domain-specific templates (e.g., dashboards, medical interfaces), the system can generate outputs that are both stylistically consistent and aligned with industry standards. Recent advances in retrieval-augmented generation \cite{lewis2020retrieval}, model context protocols \cite{hou2025model}, and LLM-generated API calls \cite{trivedi2024appworld} suggest promising directions for embedding reliable, domain-aware resources into specification-driven workflows. For the HCI community, this points to a broader implication: AI-generated UIs should not remain as isolated inspiration artifacts but evolve into structured, interoperable specifications that flow seamlessly into engineering practices. Extending SpecifyUI’s pipeline to incorporate page-level interactions, data bindings, and multi-page navigation would move the system beyond single-screen designs toward high-fidelity, executable prototypes. Such an approach can reduce handoff overhead and enable tighter collaboration between designers, developers, and AI, pushing generative systems toward production-ready co-creation.

%% file: sections/Conclusion.tex
\section{Conclusion}
Large Language Models (LLMs) create new opportunities for UI generation but leave a gap between loosely articulated design intentions and structured, implementable interfaces. To address this, we introduced SPEC, a structured intermediate representation that encodes hierarchical structure and semantic intent, and developed an LLM-based pipeline that generates and refines UIs through SPEC. Our technical evaluation showed that this pipeline improves both fidelity and controllability of generated interfaces. Building on this foundation, we implemented SpecifyUI, a prototype system that supports specification-driven generation and editing. In a user study with 16 professional UI designers, SpecifyUI significantly enhanced design quality, intent expression, and controllability compared to an industry baseline. Finally, by discussing its scalability and limitations, we highlight the broader potential of intermediate representations as a foundation for the next generation of UI generation tools.